\documentstyle[12pt,axodraw,epsfig]{article}

\begin{document}

\begin{center} 
{\Large {\bf Gluino production in some supersymmetric models at the LHC}}
\end{center}

\begin{center}
C. Brenner Mariotto and M. C. Rodriguez \\
{\it Universidade Federal do Rio Grande - FURG \\
Departamento de F\'\i sica \\
Av. It\'alia, km 8, Campus Carreiros \\
96201-900, Rio Grande, RS \\
Brazil}
\end{center}

\date{\today}


\begin{abstract}
In this article we review the mechanisms in several supersymmetric models for producing gluinos at the LHC and its potential for discovering them. We focus on the MSSM and its left-right extensions. We study in detail the strong sector of both models. Moreover, we obtain  the total cross section and differential distributions. We also make an analysis of their uncertainties, such as the gluino and squark masses, which are related to the soft SUSY breaking parameters.
\end{abstract}


\section{Introduction}

Although the Standard Model (SM) \cite{sg}, based on the gauge symmetry 
$SU(3)_{c}\otimes SU(2)_{L}\otimes U(1)_{Y}$ describes the observed properties
of charged leptons and quarks it is not the ultimate theory. 
However, the necessity to go beyond it, from the 
experimental point of view, comes at the moment only from neutrino 
data.  If neutrinos are massive then new physics beyond the SM is needed.

Supersymmetry (SUSY) or symmetry between bosons (particles with integer
spin) and fermions (particles with half-integer spin) has been
introduced in theoretical papers nearly 30 years ago~\cite{super}. 
Since that time there appeared thousands of papers.  The reason for this remarkable activity is the unique mathematical
nature of supersymmetric theories, possible solution of various
problems of the SM within its supersymmetric extentions as well as the opening perspective of
unification of all interactions in the framework of a single
theory~\cite{ssm,susy,wb}.

However supersymmetry seemed, in the early days, clearly inappropriate 
for a description of our physical world\footnote{The most physicists considered supersymmetry 
as irrelevant for ``real physics''.}, for obvious and less obvious reasons, 
which often tend to be somewhat forgotten, 
now that we got so accustomed to deal with Supersymmetric 
extensions of the Standard Model.  We recall the obstacles which seemed, long ago, to prevent
supersymmetry from possibly being a fundamental symmetry of Nature\footnote{We are grateful to P. Fayet that called our attention, and sent to us all 
these interesting information about the ``early" day of the Supersymmetric 
extensions of the Standard Model, as well as sent us all the original articles.} 
\cite{Fayet:2001xk}. 

We know that bosons and fermions should have equal masses 
in a supersymmetric theory. However, it even seemed initially
that {\it supersymmetry could not be spontaneously broken at all\,} 
which would imply that bosons and fermions be systematically degenerated 
in mass, unless of course supersymmetry-breaking terms are explicitly 
introduced ``by hand". As a result, this lead to the question:
\begin{itemize}
\item Is spontaneous supersymmetry breaking possible at all\,?

Until today, the spontaneous supersymmetry breaking remains, in general,
rather difficult to obtain.  Of course just accepting the possibility of explicit supersymmetry breaking 
without worrying too much about the origin of 
supersymmetry breaking terms,
as is frequently done now, makes things much easier
\,-- but also at the price of introducing a large number 
of arbitrary parameters, 
coefficients of these supersymmetry breaking terms.
These terms essentially serve as a parametrization 
of our ignorance about the 
true mechanism of supersymmetry breaking chosen by Nature
to make superpartners heavy. In any case such terms must have their origin 
in a spontaneous supersymmetry breaking mechanism.

However, much before getting to the Supersymmetric Standard Model,
and irrespective of the question of supersymmetry breaking, 
the crucial question,
if supersymmetry is to be relevant in particle physics, is:

\item Which bosons and fermions could be related\,?

But there seems to be no answer since known bosons
and fermions do not appear to have much in common 
\,-- except, maybe, for the photon and the neutrino. 

May be supersymmetry could act at the level of composite objects, e.g. 
as relating baryons with mesons\,?
Or should it act at a fundamental level, i.e. 
at the level of quarks and gluons\,? (But quarks are color triplets, 
and electrically charged,
while gluons transform as an $\,SU(3)\,$ color octet, 
and are electrically neutral\,!)  

In a more general way the number of (known) degrees of freedom 
is significantly larger for the fermions 
(now 90, for three families of quarks and leptons)
than for the bosons (27 \,for the gluons, the photon 
and the $\,W^\pm$ and $\,Z\,$
gauge bosons, ignoring for the moment
the spin-2 graviton, and the still-undiscovered Higgs boson).
And these fermions and bosons have very different gauge symmetry 
properties\,! This leads to the question:

\item How could one define (conserved) baryon and lepton numbers, 
in a supersymmetric theory ?

Of course nowadays we are so used to deal with spin-0 squarks and sleptons, 
carrying baryon and lepton numbers almost by definition, 
that we can hardly imagine this could once have appeared as a problem.

\end{itemize}

Supersymmetry today is the main candidate for a unified theory
beyond the SM.  Search for various manifestations of
supersymmetry in Nature is one of the main tasks of numerous
experiments at colliders.  Unfortunately, the result is negative so far. There are
no direct indications on existence of supersymmetry in particle
physics, however there are a number of theoretical and phenomenological issues that the SM fails to address 
adequately \cite{Chung:2003fi}:
\begin{itemize}
\item Unification with gravity; The point is that SUSY algebra being a generalization of
Poincar\'e algebra \cite{wb,dress,tata}
\begin{equation}
\{Q_\alpha, \bar{Q}_{\dot{\alpha}}\}=2\sigma_{\alpha,\dot{\alpha}}^{m}P_{m}.
\end{equation}
Therefore, an anticommutator of two SUSY transformations is a local coordinate translation. And
a theory which is invariant under the general coordinate transformation is General
Relativity. Thus, making SUSY local, one obtains General Relativity, or a theory of
gravity, or supergravity~\cite{sugra}.
\item Unification of Gauge Couplings; According to {\em hypothesis} of Grand Unification Theory (GUT) all gauge couplings change with energy. All known interactions are the branches of a single
interaction associated with a simple gauge group which includes the group of the SM. To reach 
this goal one has to examine how the coupling change with energy. Considerating the evolution 
of the inverse couplings, one can see that in the SM unification of the gauge
couplings is impossible. In the supersymmetric case the slopes of Renormalization Group Equation curves are changed and the results 
show that in supersymmetric model one can achieve perfect unification \cite{ABF}.
\item Hierarchy problem; The supersymmetry automatically cancels all
quadratic corrections in all orders of perturbation theory due to
the contributions of superpartners of the ordinary particles. The
contributions of the boson loops are cancelled by those of fermions
due to additional factor $(-1)$ coming from Fermi statistic. This
cancellation is true up to the SUSY breaking scale, $M_{SUSY}$,
since
\begin{equation}
 \sum_{bosons} m^2 - \sum_{fermions} m^2= M_{SUSY}^2,
\end{equation}
which  should not be very large ($\leq$ 1 TeV) to make the fine-tuning natural. 
Therefore, it provides a solution to the hierarchy 
problem by protecting the eletroweak scale from large radiative corrections 
\cite{INO82a}. However, the origin of the hierarchy is  the other part of the problem. We show below how SUSY can
explain this part as well.
\item Electroweak symmetry breaking (EWSB); The ``running" of the Higgs
masses leads to the phenomenon known as 
{\em radiative electroweak symmetry breaking}. Indeed, the mass parameters from the Higgs
potential $m_1^2$ and $m_2^2$ (or one of them) decrease while
running from the GUT scale to the scale $M_Z$ may even change the
sign. As a result for some value of the momentum $Q^2$ the potential
may acquire a nontrivial minimum. This triggers spontaneous breaking
of $SU(2)$ symmetry. The vacuum expectations of the Higgs fields
acquire nonzero values and provide masses to fermions and gauge bosons, 
and additional masses to their superpartners \cite{running}. Thus the breaking of the electroweak symmetry is not introduced by brute force as in the SM, but appears naturally from the radiative corrections.
\end{itemize}

SUSY has also made several correct predictions \cite{Chung:2003fi}:
\begin{itemize}
\item  Supersymmetry predicted in the early 1980s that the 
top quark would be heavy \cite{Ibanez:wd}, because this was a   
necessary condition for the validity of the electroweak 
symmetry breaking explanation.

\item Supersymmetric grand unified theories with a high fundamental
scale accurately predicted the present experimental value of $\sin
^{2}\theta _{W}$ before it was measured
\cite{Dimopoulos:1981yj}.

\item Supersymmetry requires a light Higgs boson to exist
\cite{Kane:1992kq}, consistent with current precision
measurements, which suggest $M_{h} < 200$ GeV \cite{lepewwg}.
\end{itemize}
Together these successes provide powerful indirect evidence that low energy SUSY is indeed part of correct description of nature.

Certainly the most popular extension of the SM is its supersymmetric counterpart 
called Minimal Supersymmetric Standard Model (MSSM) \cite{ssm,grav,mssm}. 

The first attempt to construct a phenomenological model was done in
\cite{R}, where the author tried to relate known particles together (in 
particular, the photon with a ``neutrino", and the $W^{\pm}$'s with charged 
``leptons", also related with charged Higgs bosons $H^{\pm}$),
in a $SU(2) \otimes U(1)$ electroweak theory involving two doublet Higgs
superfields now known as $H_{1}$ and $H_{2}$\footnote{Then called $S=H_{1}$ 
left-handed and $T= H_{2}^{c}$ right-handed \cite{Fayet:2001xk}.}. 
The limitations of this approach quickly led to reinterpret the
fermions of this model (which all have 1 unit of a conserved additive
R quantum number carried by the supersymmetry generator) as belonging
to a new class of particles. The ``neutrino'' ought to be considered as a really new particle,
a ``photonic neutrino'', \,a name transformed in 1977 
into {\it \,photino}; the fermionic partners of the colored gluons 
(quite distinct from the quarks) then becoming the {\it \,gluinos}, 
\,and so on.
More generally this led one to postulate the existence of new 
$\,R$-odd ``superpartners'' for all particles and consider them seriously, 
despite their rather non-conventional properties:
e.g. new bosons carrying ``fermion'' number, now known 
as {\it \,sleptons} and {\it \,squarks}, \,or Majorana fermions 
transforming as an $\,SU(3)\,$ color octet, 
which are precisely the {\it \,gluinos}, etc..
In addition the electroweak breaking must be induced by \hbox{\it a pair\,} 
of electroweak Higgs doublets, not just a single one as 
in the SM, 
which requires the existence of {\it \,charged Higgs bosons},
\,and of several neutral ones~\cite{ssm,grav}. We also want to stress that on 
reference \cite{ssm} were introduced squarks and gluinos (color octet of Majorana fermions, 
which couple to squark/quark pairs within what is now known as Supersymmetric Quantum Chromodynamics (sQCD)), 
that is the main subject of this article.

The still-hypothetical superpartners may be 
distinguished by a new quantum number called $R\,$-parity, 
\,first defined in terms of the previous $R\,$ quantum number
as $\,R_p=(-1)^R$, ~i.e. $\,+1$ 
for the ordinary particles and $\,-1\,$ for their superpartners. 
It is associated with a $\,Z_2$ remnant of the previous $\,R$-symmetry
acting continuously on gauge, lepton, quark and Higgs 
superfields as in \,\cite{ssm},
\,which must be abandoned as a continuous symmetry to allow masses for the 
gravitino~\cite{grav} and gluinos~\cite{rp}.  The conservation \,(or non-conservation)\,
of $\,R$-parity is therefore closely related with the conservation 
\,(or non-conservation)\, of baryon and lepton numbers, 
$\,B\,$ and $\,L$, ~as illustrated by the well-known formula 
reexpressing $\,R$-parity in terms of baryon and lepton numbers, 
as $\,(-1)\,^{2S} \ (-1)\,^{3B+L}$
~\cite{ff}. This may also be written as $\ (-1)^{2S} \ (-1)\,^{3\,(B-L)}\,$, 
~showing that this discrete symmetry may still be conserved 
even if baryon and lepton numbers are separately violated,
as long as their difference ($\,B-L\,$) remains 
conserved, at least modulo 2.

The finding of the basic building blocks of 
the Supersymmetric Standard Model,
whe\-ther ``minimal'' or not, allowed for the 
experimental searches for ``supersymmetric particles'', 
which started with the first searches for gluinos and photinos, 
selectrons and smuons, in the years 1978-1980,
and have been going on continuously since.
These searches often use the ``missing energy'' signature
corresponding to energy-momentum carried away by unobserved 
neutralinos~\cite{ssm,ff,ff2}.
A conserved $R$-parity also ensures the stability 
of the ``lightest supersymmetric particle'',
\,a good candidate to constitute the non-baryonic Dark Matter 
that seems to be present in the Universe.  

Massive neutrinos can also be naturally 
accommodated in $R$-parity violating supersymmetric theories,
in which neutrinos can mix with neutralinos so that they acquire small masses \cite{hall0,dreiner}. However, the
phenomenological bounds
on $B$ and/or $L$ violation \cite{dreiner,barbier}
can 
be satisfied by
imposing $B$ as a symmetry
and 
allowing the lepton number violating couplings
to be large enough to generate Majorana
neutrino masses.

However, the minimalistic extension of the MSSM is to introduce a gauge singlet superfield $\hat{N}$, this model is called ``Next Minimal Supersymmetric Standard Model" (NMSSM) \cite{dress,Fayet:1974pd}.  

It is mainly motivated by its potential to eliminate the
$\mu$ problem of the MSSM \cite{kim}, where the
origin of the
the $\mu$ parameter in the superpotential
\begin{equation}
W_{\mbox{\scriptsize MSSM}}=\mu  H_{1} H_{2}
\end{equation}
is not understood.
For phenomenological reasons it has to be of the order of the electroweak
scale,
while the ``natural" mass scale would be
of the order of the GUT or
Planck scale. This problem is evaded in the NMSSM where the
$\mu$ term in the superpotential is dynamically generated through the superpotential
\begin{equation}
W_{NMSSM} =  \lambda  \hat{H}_{1} \hat{H}_{2} \hat{N} -\frac{1}{3} k \hat{N}^{3}
+W_{MSSM}
\end{equation}
The scalar component of $\hat{N}$ is the Higgs singlet with vacuum expectation value $x$. Therefore, this 
problem is evaded in the NMSSM where the
$\mu$ term in the superpotential is dynamically generated through
$\mu = \lambda x$ with a dimensionless coupling $\lambda$.

One of the simplest extensions of the SM that 
allows to naturally explain 
the smallness of the neutrino masses (without excessively tiny Yukawa
couplings) consists in incorporating right-handed
Majorana neutrinos, and imposing a see-saw mechanism 
\footnote{For the see-saw mechanism the first paper is \cite{minkowski}} \cite{grs79,ms80a} for the neutrino
mass generation~\cite{kim,king}, it is the ``Minimal Supersymmetric Standard Model with three right 
handed neutrinos" (MSSM3RHN) \cite{tata}.

The introduction
of three families of the right-handed neutrinos $N^i$ (where $i$ is 
flavor indice) brings two new ingredients to the standard model;
one is a new scale of the Majorana masses for the right-handed neutrinos, and the
other a new matrix for Yukawa coupling constants of these new particles.  
Thus, we have two independent Yukawa matrices in the lepton
sector as in the quark sector. Therefore, this model can accommodate a see-saw mechanism,
and at the same time stabilise the hierarchy between the scale of new
physics and the electroweak (EW) scale \cite{Teixeira:2007gq}.

Other very popular ones are Left-Right symmetric theories (LRM) \cite{ps74}.  The main motivations for this model are that it gives an
explanation for the parity violation of weak interactions, provides a mechanism
(see-saw) for generating neutrino masses, and has $B-L$ as a gauge symmetry. The
model has many predictions one can directly test at a TeV-scale linear collider \cite{Huitu:1999qx}. 

The interesting and important features \cite{melfo1} of this kind of models are: 

\begin{enumerate}
\item It incorporates Left-Right (LR) symmetry \cite{ps74} 
(on top of the
quark-lepton symmetry mentioned above), which leads naturally to the
spontaneous breaking of parity and charge conjugation \cite{ps74,s79}.
 
\item It incorporates a see-saw mechanism for small neutrino masses
\cite{grs79,ms80a}.

\item It predicts the existence of magnetic monopoles \cite{p74}.

\item It leads to rare processes such as $K_L \to \mu \bar e$ through the
lepto-quark gauge bosons (with however a negligible rate for $M_{PS}
\geq 10 GeV$) \cite{ps74}.

\item In the case of single-step breaking, predicts the scale of
quark-lepton (and Left-Right) unification \cite{pss83}.
 
\item It allows naturally for $\Delta B =2$ process of $n-\bar n$ oscillations
(with however a negligible rate unless there are light diquarks in the
$TeV$ mass region)\cite{mm80}.

\item Last but not least, it allows for implementation of the
leptogenesis scenario, as suggested by 
 the see-saw mechanism \cite{fy86}. 
\end{enumerate}

On the technical side, the left-right symmetric model has a problem similar to
that in the SM: the masses of the fundamental Higgs scalars diverge
quadratically. As in the SM, the Supersymmetric Left-Right model (SUSYLR) can be used to stabilize the scalar
masses and cure this hierarchy problem.

On the literature there are two different SUSYLR models. They differ in their
$SU(2)_{R}$ breaking fields: one uses $SU(2)_{R}$ triplets \cite{susylr} (SUSYLRT) and the
other $SU(2)_{R}$ doublets \cite{doublet} (SUSYLRD). 
 
Another,
 maybe more important {\it raison d'etre} for SUSYLR is the fact that they lead naturally to R-parity conservation  \cite{melfo2}.
Namely, Left-Right models contain a $B-L$ gauge symmetry, which allows
for this possibility \cite{m86}. All that is  needed is that one uses
a version  of the theory that incorporates a see-saw mechanism \cite{grs79,ms80a}
at the renormalizable level.

As we said before, the supersymmetric particles have not yet been detected in the present machines such
as HERA and Tevatron. By another hand there are many interesting supersymmetric models in the literature as we 
shown above. Therefore, if 
SUSY is detected in the Large Hadron Collider (LHC), 
one of the next steps
will be to discriminate among the different SM extensions, scenarios and
also to find the mass spectrum of the different particles (which can be obtained
theoretically in different scenarios). 

To discriminate among the several 
possibilities, it is important to make predictions for different 
observables and confront these predictions with the forthcoming experimental
data. One important process which could be measured at the LHC is the gluino 
production.

The $\,R$-symmetry transformations act {\it \,chirally\,} 
on gluinos, so that an unbroken $\,R$-invariance
would require them to remain massless,
even after a spontaneous breaking of the supersymmetry\,!
In the early days it was very difficult to obtain large masses for gluinos, 
since: \ 
{\bf i)} \ no direct gluino mass term was present in the Lagrangian density; 
and \ 
{\bf ii)} \ no such term may be generated spontaneously, at the tree 
approximation, since gluino couplings involve {\it colored\,} 
\hbox{spin-0} fields.

On this case, gluino remain massless, and we would then expect the existence of relatively light
``$R$-hadrons''~\cite{ff,ff2} made of quarks, antiquarks and gluinos, 
which have not been observed. We know today that gluinos, 
if they do exist, should be rather heavy, requiring a significant 
breaking of the continuous $\,R$-invariance,
in addition to the necessary breaking of the supersymmetry.

A third reason for abandoning the continuous $\,R$-symmetry 
could now be the non-observation at LEP
of a charged {\it wino} \,-- also called {\it \,chargino\,} --\,
lighter than the $\,W^\pm$, \,that would exist 
in the case of a continuous $\,U(1)\,$ $\,R$-invariance~\cite{ssm,R}. 
The just-discovered $\,\tau^-\,$ particle could tentatively be considered, 
in 1976, as a possible light wino/chargino candidate, 
before getting clearly identified as a sequential heavy lepton.

The {\bf gluino masses} result directly from {\bf \underline{supergravity}}, through 
$m_{3/2}$, which leads one to abandon continuous R for R-parity 
as already observed in 1977~\cite{grav}. Another way does not use SUGRA but generates $m_{gluino}$ radiatively,
using messengers quarks sensitive to the source of SUSY-breaking~\cite{glu}, on this case the gluino masses are generated by radiative corrections involving a new sector of quarks sensitive 
to the source of supersymmetry breaking,
that would now be called  ``messenger quarks'' \footnote{Remember that the see-saw mechanism
 \cite{minkowski} did not attract attention at 
the time, however the article \cite{glu} on gluino masses discusses a see-saw mechanism for gluinos.}.

Today, the  
gluinos are expected to be one of the most massive sparticles which 
constitute the Minimal Supersymmetric Standard Model (MSSM), and therefore 
their production is only feasible at a very energetic machine such as the 
LHC. Being the fermion partners of the gluons, their 
role and interactions are directly related with the properties of the 
supersymmetric QCD (sQCD).

The aim of this paper is twofold. The first one is to study the strong sector of some supersymmetric models and to show 
explicit that the Feynmann rules for the gluino production are the same in all supersymmetric extensions of the 
Standard Model here considered. After that, as the second aim of this article, we show predictions for 
the gluino production on these models at the LHC, 
for various SPS benchmark points. 
The outline of the paper is the following. In sections \ref{sec:ma} and 
\ref{sec:susylr}, we obtain the relevant Feynman rules of the strong sector from
the MSSM and SUSYLR models, respectivelly. Moreover, we see that the
Feynman rules are indeed the same in these models. In section
\ref{spspoints} we consider the different scenarious for the relevant SUSY parameters, 
which lead to different gluinos and squark masses. In section 
\ref{gluinoprod} we consider gluino production in the studied models 
and present the relevant expressions, which are used to obtain the numerical results 
in section \ref{results}. 
Conclusions are summarized in section \ref{sec:concl}.

\section{Minimal Supersymmetric Standard Model (MSSM).}

\label{sec:ma}

In the MSSM \cite{ssm,grav,mssm}, the gauge group is $SU(3)_{C}\otimes
SU(2)_{L}\otimes U(1)_{Y}$. The particle content of this model consists in associate to every
known quark and lepton a new scalar superpartner to form a chiral supermultiplet. Similarly, we
group a gauge fermion (gaugino) with each of the gauge bosons of the standard model to form a
vector multiplet. In the scalar sector, we need to introduce two Higgs scalars and also their
supersymmetric partners known as Higgsinos (Our notation 
\footnote{The particle content and the Lagrangian of this model.} is given at \cite{cmmc}). We also need to impose a new global $U(1)$ invariance 
usually called $R$-invariance, to get interactions that conserve both lepton and
baryon number (invariance). On this section we will derive the Feynman rules of the strong  
sector of this model.

\subsection{Interaction from ${\cal L}_{Gauge}$}

We can rewrite ${\cal L}_{Gauge}$, see \cite{dress,tata}, in the following way
\begin{eqnarray}
{\cal L}_{\mbox{Gauge}}
&=& {\cal L}_{cin}+{\cal L}_{gaugino}+{\cal L}^{gauge}_{D}.
\end{eqnarray}
where
\begin{eqnarray}
{\cal L}_{cin}&=&{\cal L}^{SU(3)}_{cin}+{\cal L}^{SU(2)}_{cin}+{\cal L}^{U(1)}_{cin}, \nonumber \\
{\cal L}_{gaugino}&=&{\cal L}^{SU(3)}_{gaugino}+{\cal L}^{SU(2)}_{gaugino}+{\cal L}^{U(1)}_{gaugino}, \nonumber \\
 {\cal L}^{gauge}_{D}&=&{\cal L}^{SU(3)}_{D}+{\cal L}^{SU(2)}_{D}+{\cal L}^{U(1)}_{D},
\end{eqnarray}
the first part is given by:
\begin{equation}
{\cal L}^{SU(3)}_{cin}=- \frac{1}{4}G^{a}_{mn}G^{amn},
\label{usualqcd}
\end{equation}
with
\begin{equation}
G^{a}_{mn}= \partial_{m}g^{a}_{n}-\partial_{n}g^{a}_{m}-g_{s}f^{abc}g^{b}_{m}
g^{c}_{n}, 
\label{usualqcd1}
\end{equation}
$g_{s}$ is the strong coupling constant and $f^{abc}$ are totally antisymmetric structure constant of 
the $SU(3)$ group . The second term can be rewriten as
\begin{equation}
{\cal L}^{SU(3)}_{gaugino}=- \imath \overline{\lambda^{a}_{C}}\bar{\sigma}^{m}{\cal D}_{m}\lambda^{a}_{C},
\label{gaugino1}
\end{equation}
where
\begin{equation}
{\cal D}_{m}\lambda^{a}_{C}= \partial_{m}\lambda^{a}_{C}-g_{s}f^{abc}\lambda^{a}_{C}g^{c}_{m}.
\label{gaugino2}
\end{equation}
The last term is given by
\begin{equation}
{\cal L}^{SU(3)}_{D}= \frac{1}{2}D^{a}_{C}D^{a}_{C}.
\end{equation}

\subsubsection{Gluon Self Interaction}

These interactions are derived from Eqs.(\ref {usualqcd},\ref{usualqcd1}) which are the same as the usual QCD. Therefore, 
on this case, we get the same Feynman rules as those from the QCD.

\subsubsection{Gluino--Gluino--Gluon Interaction }

This interaction is got from Eqs.(\ref{gaugino1}, \ref{gaugino2}) combining both equations to obtain
\begin{equation}
{\cal L}_{gaugino}={\cal L}_{cin}+{\cal L}_{\tilde{g}\tilde{g}g},
\end{equation}
where
\begin{eqnarray}
{\cal L}_{cin}&=& \imath ( \partial_{m} \overline{\lambda^{a}}_{C}) \bar{\sigma}^{m}\lambda^{a}_{C}, \nonumber \\ 
{\cal L}_{\tilde{g}\tilde{g}g}&=& \imath g_{s}f^{abc}  \overline{\lambda^{a}}_{C} \bar{\sigma}^{m}\lambda^{b}_{C}g^{c}_{m},
\end{eqnarray}
the first term gives the cinetic term to gluino, while the last one provides the gluino-gluino-gluon interaction.

Considerating the four-component Majorana spinor for the gluino, given by
\begin{equation}
\Psi(\tilde{g}^{a}) \;=\; \left( \begin{array}{r} - \imath \lambda^{a}_{C}(x) \\
                    \imath \overline{\lambda^{a}}_{C}(x) \end{array} \right) \,\ .,
                        \label{gluino spinor}
\end{equation}
we can rewrite 
${\cal L}_{\tilde{g}\tilde{g}g}$ in the following way
\begin{equation}
{\cal L}_{\tilde{g}\tilde{g}g}= \frac{\imath}{2} g_{s}f^{bac} \bar{\Psi}(\tilde{g}^{a})\gamma^{m}\Psi(\tilde{g}^{b})g^{c}_{m}.
\label{regrafeynman1}
\end{equation} 
Owing to the Majorana nature of the gluino one must multiply 
by 2 to obtain the Feynman rule (or add the graph with 
$\tilde{g} \leftrightarrow \bar{\tilde{g}}$)!

\begin{figure}[h]
\begin{center}
\begin{picture}(150,100)(0,0)
\SetWidth{1.2}
\Line(10,10)(50,50)
\Photon(10,10)(50,50){4}{5}
\Text(5,10)[r]{$\tilde{g}^{a}$}
\Line(10,90)(50,50)
\Photon(10,90)(50,50){4}{5}
\Text(5,90)[r]{$\tilde{g}^{b}$}
\Vertex(50,50){1.5}
\Text(60,40)[]{$g^{c}_{m}$}
\Gluon(100,50)(50,50){4}{5}
\Text(130,50)[]{$-g_{s}f^{bac}\gamma^{m}$}
\end{picture}
\end{center}
\caption{Feynman rule for the vertice Gluino-Gluino-Gluon at the MSSM.}
\label{fig1}
\end{figure}
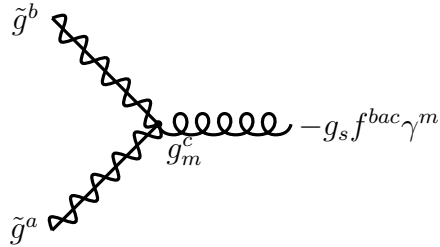

The equation above induce the following Feynman rule, for the vertice gluino-gluino-gluon, given 
at Fig.(\ref{fig1}) and
it is the same result as presented at \cite{dress,tata,mssm,kraml}.

\subsection{Interaction from ${\cal L}_{\mbox{Quarks}}$}

The interaction of the strong sector are obtained from the following Lagrangians
\begin{eqnarray}
{\cal L}_{qqg}&=&g_{s} \bar{Q} \bar{\sigma}_{m}T^{a}Qg^{am}+
g_{s} \overline{u^{c}} \bar{\sigma}_{m}\bar{T}^{a}u^{c}g^{am}+
g_{s} \overline{d^{c}} \bar{\sigma}_{m}\bar{T}^{a}d^{c}g^{am} \nonumber \\
{\cal L}_{ \tilde{q} \tilde{q}g}&=&\imath g_{s} \left( 
\bar{\tilde{Q}}T^{a}\partial_{m}\tilde{Q}-
\tilde{Q}T^{a}\partial_{m}\bar{\tilde{Q}} \right) g^{am}+
\imath g_{s} \left( 
\overline{\tilde{u}^{c}}\bar{T}^{a}\partial_{m}\tilde{u}^{c}-
\tilde{u}^{c}\bar{T}^{a}\partial_{m}\overline{\tilde{u}^{c}} \right) g^{am}
\nonumber \\ &+&
\imath g_{s} \left( 
\overline{\tilde{d}^{c}}\bar{T}^{a}\partial_{m}\tilde{d}^{c}-
\tilde{d}^{c}\bar{T}^{a}\partial_{m}\overline{\tilde{d}^{c}} \right) g^{am}, \nonumber \\
{\cal L}_{q \tilde{q} \tilde{g}}&=&- \imath \sqrt{2}g_{s} \left(
\bar{Q}T^{a}\tilde{Q}\overline{\lambda^{a}_{C}}-
\bar{\tilde{Q}}T^{a}Q\lambda^{a}_{C} \right) -
\imath \sqrt{2}g_{s} \left(
\overline{u^{c}}\bar{T}^{a}\tilde{u}^{c}\overline{\lambda^{a}_{C}}-
\overline{\tilde{u}^{c}}\bar{T}^{a}u^{c}\lambda^{a}_{C} \right) 
\nonumber \\
 &-&
\imath \sqrt{2}g_{s} \left(
\overline{d^{c}}\bar{T}^{a}\tilde{d}^{c}\overline{\lambda^{a}_{C}}-
\overline{\tilde{d}^{c}}\bar{T}^{a}d^{c}\lambda^{a}_{C} \right), 
\nonumber \\
{\cal L}_{ \tilde{q} \tilde{q}gg}&=&-g_{s}^{2}\bar{\tilde{Q}}T^{a}T^{b}\tilde{Q}g^{a}_{m}g^{bm}
-g_{s}^{2}\overline{\tilde{u}^{c}}\bar{T}^{a}\bar{T}^{b}\tilde{u}^{c}g^{a}_{m}g^{bm}-
g_{s}^{2}\overline{\tilde{d}^{c}}\bar{T}^{a}\bar{T}^{b}\tilde{d}^{c}g^{a}_{m}g^{bm}.
\label{quarksint}
\end{eqnarray}
Where $T^{a}_{rs}$ are the color triplet generators, then one must use 
\begin{equation}
\bar{T}^{a}_{rs}=- T^{* a}_{rs}=-T^{a}_{sr},
\label{antitripletrepresentation}
\end{equation}
for the color anti-triplet generators.

\subsubsection{Quark--Quark--Gluon Interaction}

\begin{figure}[h]
\begin{center}
\begin{picture}(150,100)(0,0)
\SetWidth{1.2}
\Line(10,10)(50,50)
\Text(5,10)[r]{$q_{r}$}
\Line(10,90)(50,50)
\Text(5,90)[r]{$q_{s}$}
\Vertex(50,50){1.5}
\Text(60,40)[]{$g^{a}_{m}$}
\Gluon(100,50)(50,50){4}{5}
\Text(130,50)[]{$- \imath g_{s} T^{a}_{rs} \gamma^{m}$}
\end{picture}
\end{center}
\caption{Feynman rule for the vertice Quark-Quark-Gluon at the MSSM.}
\label{fig2}
\end{figure}
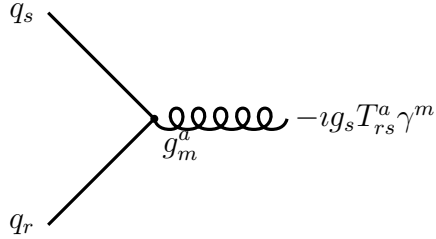

This interaction comes from the first Lagrangian given at Eq.(\ref{quarksint}),  and can be rewritten as
\begin{equation}
{\cal L}_{qqg}=g_{s}( \bar{u}_{r}\bar{\sigma}^{m}T^{a}_{rs}u_{s}+  \bar{d}_{r}\bar{\sigma}^{m}T^{a}_{rs}d_{s}+
\overline{u^{c}}_{r}\bar{\sigma}^{m}\bar{T}^{a}_{rs}u^{c}_{s}+  \overline{d^{c}}_{r}\bar{\sigma}^{m}\bar{T}^{a}_{rs}d^{c}_{s})g^{a}_{m},
\label{qqgmssm}
\end{equation}
$u$ and $d$ are color triplets while $u^{c}$ and $d^{c}$ are color anti-triplets. We must recall that $r$ and $s$ are color indices.  Using 
Eq.(\ref{antitripletrepresentation}) we can rewrite Eq.(\ref{qqgmssm}) as
\begin{equation}
{\cal L}_{qqg}=g_{s}( \bar{u}_{r}\bar{\sigma}^{m}T^{a}_{rs}u_{s}+  \bar{d}_{r}\bar{\sigma}^{m}T^{a}_{rs}d_{s}+
u^{c}_{r}\sigma^{m}T^{a}_{rs}\overline{u^{c}}_{s}+d^{c}_{r}\sigma^{m}T^{a}_{rs}\overline{d^{c}}_{s})g^{a}_{m}. 
\end{equation}
Now, if we take into account the four-component Dirac spinor of the quarks ($q=u,d$), given by 
\begin{equation}
\Psi(q) \;=\; \left( \begin{array}{r} q_{L}(x) \\
                    \overline{q^{c}}_{L}(x) \end{array} \right) \,\ , 
                        \label{quarks spinors}
\end{equation}
we obtain the Feynman rule
\begin{equation}
{\cal L}_{qqg}=-g_{s}\sum_{q=u,d} \bar{\Psi}(q_{r})\gamma^{m}T^{a}_{rs}\Psi(q_{s})g^{a}_{m},
\label{regrafeynman2}
\end{equation}
for the vertice $qqg$ drawn at Fig.(\ref{fig2})
this results again, as expected, agree with the results given at \cite{dress,tata,mssm,kraml}.

The gluino-gluon interaction is like the quark-gluon interaction, compare Fig.(\ref{fig1}) with Fig.(\ref{fig2}) \footnote{This fact as 
first noticed at \cite{ssm}.} and this  fact gives the idea of ``$R$-hadrons'' presented at~\cite{ff,ff2}, mentioned in our introduction.

\subsubsection{Squark--Squark--Gluon Interaction}

This interaction is obtained from the second line given at Eq.(\ref{quarksint}). This term can be rewritten in the following way
\begin{eqnarray}
{\cal L}_{\tilde{q}\tilde{q}g}&=& \imath g_{s}(  \bar{\tilde{u}}_{r}T^{a}_{rs}(\partial_{m}\tilde{u}_{s})-
(\partial_{m}\overline{\tilde{u}}_{r}) T^{a}_{rs}\tilde{u}_{s}+  \bar{\tilde{d}}_{r}T^{a}_{rs}(\partial_{m}\tilde{d}_{s})-
(\partial_{m}\overline{\tilde{d}}_{r}) T^{a}_{rs}\tilde{d}_{s} \nonumber \\ &+&
\overline{\tilde{u}^{c}}_{r}\bar{T}^{a}_{rs}(\partial_{m}\tilde{u}^{c}_{s})-
(\partial_{m}\overline{\tilde{u}^{c}}_{r}) \bar{T}^{a}_{rs}\tilde{u}^{c}_{s}+  
\overline{\tilde{d}^{c}}_{r}\bar{T}^{a}_{rs}(\partial_{m}\tilde{d}^{c}_{s})-
(\partial_{m}\overline{\tilde{d}^{c}}_{r}) \bar{T}^{a}_{rs}\tilde{d}^{c}_{s})g^{am}. \nonumber \\
\label{sqsqgmssm}
\end{eqnarray}
Using the following identity
\begin{equation}
  A {\!\stackrel{\leftrightarrow}{\partial}_{m}\!} B \equiv A\,(\partial_{m} B) - (\partial_{m} A)\,B 
\end{equation}
we can rewrite our Lagrangian in the following simple way
\begin{equation}
{\cal L}_{\tilde{q}\tilde{q}g}= \imath g_{s} \sum_{q=u,d} (
\tilde{q}_{Lr}^{*}T^{a}_{rs}\,  {\!\stackrel{\leftrightarrow}{\partial}_{m}\!} \, \tilde{q}_{Ls}-
\tilde{q}_{Rr}^{*}T^{a}_{rs}\,  {\!\stackrel{\leftrightarrow}{\partial}_{m}\!}\,  \tilde{q}_{Rs} )g^{am}\,.
\end{equation}
Note the relative minus sign between the terms with $\tilde{q}_{L}$ and $\tilde{q}_{R}$: 
This is due to the facts that $\tilde{q}_{R}$ are colour anti--triplets and  
the anti--colour generator given at Eq.(\ref{antitripletrepresentation}). Here, we use a similar notation 
as given at \cite{mssm}, it means that $\tilde{q}^{*}$ creates the scalar quark $\tilde{q}$, while $\tilde{q}$ 
destroys the scalar quark $\tilde{q}$.

Including the generalization to six flavors, we can write
\begin{equation}
{\cal L}_{\tilde{q}\tilde{q}g}= \imath g_{s} \sum_{q=u,d} \sum_{p=1}^{6} (
\tilde{q}_{Lpr}^{*}T^{a}_{rs}\,  {\!\stackrel{\leftrightarrow}{\partial}_{m}\!} \, \tilde{q}_{Lps}-
\tilde{q}_{Rpr}^{*}T^{a}_{rs}\,  {\!\stackrel{\leftrightarrow}{\partial}_{m}\!}\,  \tilde{q}_{Rps} )g^{am}
\label{squarksflavor}
\end{equation}

\begin{figure}[h]
\begin{center}
\begin{picture}(150,100)(0,0)
\SetWidth{1.2}
\DashLine(10,10)(50,50){4.5}
\Text(5,10)[r]{$\tilde{q}_{r}$}
\DashLine(50,50)(10,90){4.5}
\Text(5,90)[r]{$\tilde{q}_{s}$}
\Vertex(50,50){1.5}
\Text(60,40)[]{$g^{a}_{m}$}
\Gluon(100,50)(50,50){4}{5}
\Text(150,50)[]{$- \imath g_{s}T^{a}_{rs}(k_i^{} + k_j^{})^{m}$}
\end{picture}
\end{center}
\caption{Feynman rule for the vertice Squark-Squark-Gluon at the MSSM.}
\label{fig3}
\end{figure}
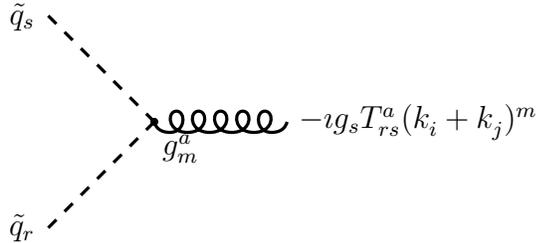

The corresponding Feynman rule we obtain from
\begin{equation}
  \tilde{q}_{j}^{*}\,  {\!\stackrel{\leftrightarrow}{\partial}^{m}\!}\,  \tilde{q}_{i} = \imath \,\ (k_i^{} + k_j^{})^{m}
\label{regrafeynman3}
\end{equation}
where $k_i^{}$ and $k_j^{}$ are the four--momenta of $\tilde{q}_{i}$ and $\tilde{q}_{j}$
in direction of the charge flow. This relations give us the following Feynman rules given at Fig.(\ref{fig3}) and we conclude 
that
this results, as expected, agree with the references  \cite{dress,tata,mssm,kraml}.

\subsubsection{Squark-Squark-Gluon-Gluon Interaction}
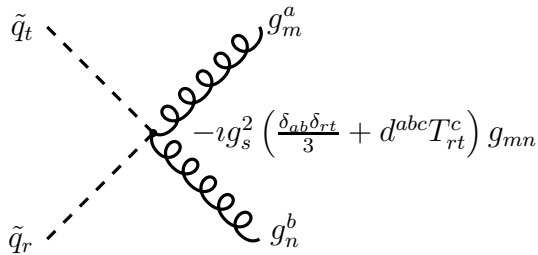
\begin{figure}[h]
\begin{center}
\begin{picture}(150,100)(0,0)
\SetWidth{1.2}
\DashLine(10,10)(50,50){4.5}
\Text(5,10)[r]{$\tilde{q}_{r}$}
\DashLine(10,90)(50,50){4.5}
\Text(5,90)[r]{$\tilde{q}_{t}$}
\Vertex(50,50){1.5}
\Gluon(90,90)(50,50){4}{5}
\Text(100,90)[b]{$g^{a}_{m}$}
\Gluon(90,10)(50,50){4}{5}
\Text(100,10)[b]{$g^{b}_{n}$}
\Text(130,50)[]{$- \imath g^{2}_{s} \left( \frac{\delta_{ab}\delta_{rt}}{3}+d^{abc}T^{c}_{rt} \right)g_{mn}$}
\end{picture}
\end{center}
\caption{Feynman rule to Squark-Squark-Gluon-Gluon vertice at MSSM.}
\label{fig4}
\end{figure}

This interaction comes from the last line given at Eq.(\ref{quarksint}), which can be written as
\begin{eqnarray}
{\cal L}_{\tilde{q}\tilde{q}gg}&=&-g^{2}_{s}(
\bar{\tilde{u}}_{r}T^{a}_{rs}T^{b}_{st}\tilde{u}_{t}+
\bar{\tilde{d}}_{r}T^{a}_{rs}T^{b}_{st}\tilde{d}_{t}+
\overline{\tilde{u}^{c}}_{r}\bar{T}^{a}_{rs}\bar{T}^{b}_{st}\tilde{u}^{c}_{t}+
\overline{\tilde{d}^{c}}_{r}\bar{T}^{a}_{rs}\bar{T}^{b}_{st}\tilde{d}^{c}_{t})g^{a}_{m}g^{bm}. \nonumber \\
\label{sgsgggmssm}
\end{eqnarray}
By using the following formula valid for $SU(3)$ generators
\begin{equation}
T^{a}_{rs}T^{b}_{st}= \frac{1}{6}\delta_{ab}\delta_{rt}+ \frac{1}{2}(d^{abc}+ \imath f^{abc})T^{c}_{rt},
\end{equation}
it allows us to rewrite our Lagrangian in the following way\footnote{Including the generalization to six flavors, see Eq.(\ref{squarksflavor}).}
\begin{eqnarray}
{\cal L}_{\tilde{q}\tilde{q}gg}&=&- \frac{g^{2}_{s}}{6} \sum_{q=u,d} \tilde{q}^{*}_{r}\tilde{q}_{r}g^{a}_{m}g^{am}-
g^{2}_{s}(d^{abc}+ \imath f^{abc})  \sum_{q=u,d} \tilde{q}^{*}_{r}T^{c}_{rt}\tilde{q}_{t}g^{a}_{m}g^{bm},
\label{regrafeynman4}
\end{eqnarray}
however $f^{abc}g^{a}_{m}g^{bm}=0$ because $f^{abc}$ is totally antisymmetric structure constant of 
the group $SU(3)$ while $g^{a}_{m}g^{bm}$ 
is symmetric ones. The Feynman rule is drawn at Fig.(\ref{fig4}) and it is in agreement with \cite{dress,tata,mssm,kraml} .

\subsubsection{Gluino-Quark-Squark Interaction}
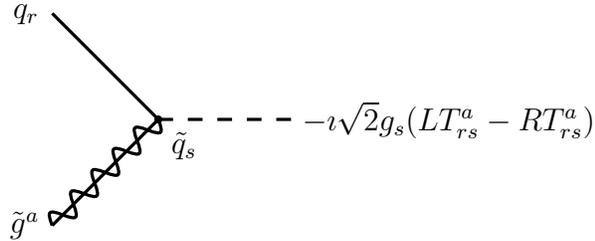
\begin{figure}[h]
\begin{center}
\begin{picture}(150,100)(0,0)
\SetWidth{1.2}
\Line(10,10)(50,50)
\Photon(10,10)(50,50){4}{5}
\Text(5,10)[r]{$\tilde{g}^{a}$}
\Line(10,90)(50,50)
\Text(5,90)[r]{$q_{r}$}
\Vertex(50,50){1.5}
\Text(60,40)[]{$\tilde{q}_{s}$}
\DashLine(100,50)(50,50){6}
\Text(160,50)[]{$- \imath \sqrt{2}g_{s}(LT^{a}_{rs}-RT^{a}_{rs})$}
\end{picture}
\end{center}
\caption{Feynman rule for the vertice Gluino-Quark-Squarks at the MSSM.}
\label{fig5}
\end{figure}

This interaction is described by the third line at Eq.(\ref{quarksint}), writing this term as
\begin{eqnarray}
{\cal L}_{\tilde{q}qg}&=&- \sqrt{2}\imath g_{s}( \bar{u}_{r}T^{a}_{rs}\tilde{u}_{s}\overline{\lambda^{a}}_{C}-
\overline{\tilde{u}}_{r}T^{a}_{rs}u_{s}\lambda^{a}_{C}+  \bar{d}_{r}T^{a}_{rs}\tilde{d}_{s}\overline{\lambda^{a}}_{C}-
\overline{\tilde{d}}_{r}T^{a}_{rs}d_{s}\lambda^{a}_{C} \nonumber \\ &+&
 \overline{u^{c}}_{r}\bar{T}^{a}_{rs}\tilde{u}^{c}_{s}\overline{\lambda^{a}}_{C}-
\overline{\tilde{u}^{c}}_{r}\bar{T}^{a}_{rs}u^{c}_{s}\lambda^{a}_{C}+
\overline{d^{c}}_{r}\bar{T}^{a}_{rs}\tilde{d}^{c}_{s}\overline{\lambda^{a}}_{C}-
\overline{\tilde{d}^{c}}_{r}\bar{T}^{a}_{rs}d^{c}_{s}\lambda^{a}_{C}).
\label{sgqsq}
\end{eqnarray}
Using the Eqs.(\ref{quarks spinors},\ref{gluino spinor}) and the usual chiral projectors
\begin{eqnarray}
L = \frac{1}{2} \left(1+ \gamma_{5} \right), \,\
R = \frac{1}{2} \left(1- \gamma_{5} \right) \,\ ,
\label{projectors}
\end{eqnarray}
we can rewrite our Lagrangian in the following way
\begin{eqnarray}
{\cal L}_{\tilde{q}qg}&=&- \sqrt{2}g_{s} \sum_{q=u,d}( 
\bar{\Psi}( \tilde{g}^{a})L\Psi(q_{r})T^{a}_{rs}\tilde{q}^{*}_{sL}+
 \bar{\Psi}( q_{r})RT^{a}_{rs}\Psi( \tilde{g}^{a})\tilde{q}_{sL}-
 \bar{\Psi}( q_{r})LT^{a}_{rs}\Psi( \tilde{g}^{a})\tilde{q}_{sL} \nonumber \\ &-&
\bar{\Psi}( \tilde{g}^{a})R\Psi(q_{r})T^{a}_{rs}\tilde{q}^{*}_{sR}).
\label{regrafeynman5}
\end{eqnarray}
this equation give us the following Feynman rule, given at Fig.(\ref{fig5}), and
again it is in concordance with \cite{dress,tata,mssm,kraml}.

Before considerating the left-right models, we want to say that the color sector of the interesting 
models NMSSM and MSSM3RHN are the same as in the MSSM model, therefore the results 
presented above are still hold on these models.

\section{Supersymmetric Left-Right Model (SUSYLR)}

\label{sec:susylr}

The supersymmetric extension of left-right models is
based on the gauge group $SU(3)_{C}\otimes SU(2)_{L}\otimes SU(2)_{R}\otimes
U(1)_{B-L}$.  On the literature, as we said at introduction, there are two different SUSYLR models. They differ in their
$SU(2)_{R}$ breaking fields: one uses $SU(2)_{R}$ triplets \cite{susylr} (SUSYLRT) and the
other $SU(2)_{R}$ doublets \cite{doublet} (SUSYLRD). Some details of both models are described at
\cite{cmmc}. SUSYLR models have the additional
appealing characteristics of having automatic R-parity conservation.

In this article, we are interested in studying only the strong sector. 
As this sector is the same in both models, SUSYLRT and SUSYLRD, the results we 
are presenting in this section hold in both models.

\subsection{Interaction from ${\cal L}_{Gauge}$}

We can rewrite ${\cal L}_{Gauge}$, as done in the MSSM case, in the following way
\begin{eqnarray}
{\cal L}_{\mbox{Gauge}}
&=& {\cal L}_{cin}+{\cal L}_{gaugino}+{\cal L}^{gauge}_{D}.
\end{eqnarray}
where
\begin{eqnarray}
{\cal L}_{cin}&=&{\cal L}^{SU(3)}_{cin}+{\cal L}^{SU(2)_{L}}_{cin}+{\cal L}^{SU(2)_{R}}_{cin}+{\cal L}^{U(1)}_{cin}, \nonumber \\
{\cal L}_{gaugino}&=&{\cal L}^{SU(3)}_{gaugino}+{\cal L}^{SU(2)_{L}}_{gaugino}+{\cal L}^{SU(2)_{R}}_{gaugino}+
{\cal L}^{U(1)}_{gaugino}, \nonumber \\
{\cal L}^{gauge}_{D}&=&{\cal L}^{SU(3)}_{D}+{\cal L}^{SU(2)_{L}}_{D}+{\cal L}^{SU(2)_{R}}_{D}+
{\cal L}^{U(1)}_{D}, 
\end{eqnarray}
the first part is given by:
\begin{equation}
{\cal L}^{SU(3)}_{cin}=- \frac{1}{4}G^{a}_{mn}G^{amn},
\label{usualqcdlr}
\end{equation}
with $G^{a}_{mn}$ is given by Eq.(\ref{usualqcd1}) 
\begin{equation}
{\cal L}^{SU(3)}_{gaugino}=- \imath \overline{\lambda^{a}_{C}}\bar{\sigma}^{m}{\cal D}_{m}\lambda^{a}_{C},
\label{gaugino1lr}
\end{equation}
while ${\cal D}^{C}_{n}\lambda^{a}_{C}$ is defined at Eq.(\ref{gaugino2}). The last term is given by
\begin{equation}
{\cal L}^{SU(3)}_{D}= \frac{1}{2}D^{a}_{C}D^{a}_{C}.
\end{equation}

The Eqs.(\ref{usualqcdlr},\ref{gaugino1lr}) are equal from the Lagrangian of the MSSM, given at Eqs.(\ref{usualqcd},\ref{gaugino1}). 
Therefore the Feynman rules are the same as in the MSSM case. The Feynman rule of the interaction Gluino-Gluino-Gluon is 
given at Eq.(\ref{regrafeynman1}).

\subsection{Interaction from ${\cal L}_{\mbox{Quarks}}$}

In terms of the doublets the Lagrangian of the strong sector can be rewritten as
\begin{eqnarray}
{\cal L}_{qqg}&=&g_{s} \bar{Q} \bar{\sigma}_{m}T^{a}Qg^{am}+
g_{s} \bar{Q}^{c} \bar{\sigma}_{m}\bar{T}^{a}Q^{c}g^{am} \nonumber \\
{\cal L}_{ \tilde{q} \tilde{q}g}&=& \imath g_{s} \left( 
\bar{\tilde{Q}}T^{a}\partial_{m}\tilde{Q}-
\tilde{Q}T^{a}\partial_{m}\bar{\tilde{Q}} \right) g^{am}+
\imath g_{s} \left( 
\bar{\tilde{Q}}^{c}\bar{T}^{a}\partial_{m}\tilde{Q}^{c}-
\tilde{Q}^{c}\bar{T}^{a}\partial_{m}\bar{\tilde{Q}}^{c} \right) g^{am}, \nonumber \\
{\cal L}_{q \tilde{q} \tilde{g}}&=&- \imath \sqrt{2}g_{s} \left(
\bar{Q}T^{a}\tilde{Q}\bar{\tilde{g}}^{a}-
\bar{\tilde{Q}}T^{a}Q\tilde{g}^{a} \right) -
\imath \sqrt{2}g_{s} \left(
\bar{Q}^{c}\bar{T}^{a}\tilde{Q}^{c}\bar{\tilde{g}}^{a}-
\bar{\tilde{Q}}^{c}\bar{T}^{a}Q^{c}\tilde{g}^{a} \right) , \nonumber \\
{\cal L}_{ \tilde{q} \tilde{q}gg}&=&g_{s}^{2}\bar{\tilde{Q}}T^{a}T^{b}\tilde{Q}g^{a}_{m}g^{bm}
-g_{s}^{2}\bar{\tilde{Q}}^{c}\bar{T}^{a}\bar{T}^{b}\tilde{Q}^{c}g^{a}_{m}g^{bm}. \nonumber \\
\label{quarkinteslr}
\end{eqnarray}

\subsubsection{Quark--Quark--Gluon Interaction }

This interaction is given by the first line given at Eq.(\ref{quarkinteslr}). Using the doublets 
we can write our Lagrangian in the following way
\begin{equation}
{\cal L}_{qqg}=g_{s}( \bar{u}_{r}\bar{\sigma}^{m}T^{a}_{rs}u_{s}+  \bar{d}_{r}\bar{\sigma}^{m}T^{a}_{rs}d_{s}+
\overline{u^{c}}_{r}\bar{\sigma}^{m}\bar{T}^{a}_{rs}u^{c}_{s}+ 
\overline{d^{c}}_{r}\bar{\sigma}^{m}\bar{T}^{a}_{rs}d^{c}_{s})
g^{a}_{m},
\end{equation}
which is the same as of the MSSM, see Eq.(\ref{qqgmssm}), therefore the Feynman rule is again given by Eq.(\ref{regrafeynman2}).

\subsubsection{Squark--Squark--Gluon Interaction}

This interaction comes from the second line at Eq.(\ref{quarkinteslr}), and can be rewritten as
\begin{eqnarray}
{\cal L}_{\tilde{q}\tilde{q}g}&=& \imath g_{s}(  \bar{\tilde{u}}_{r}T^{a}_{rs}(\partial_{m}\tilde{u}_{s})-
(\partial_{m}\overline{\tilde{u}}_{r}) T^{a}_{rs}\tilde{u}_{s}+  \bar{\tilde{d}}_{r}T^{a}_{rs}(\partial_{m}\tilde{d}_{s})-
(\partial_{m}\overline{\tilde{d}}_{r}) T^{a}_{rs}\tilde{d}_{s} \nonumber \\ &+&
\overline{\tilde{u}^{c}}_{r}\bar{T}^{a}_{rs}(\partial_{m}\tilde{u}^{c}_{s})-
(\partial_{m}\overline{\tilde{u}^{c}}_{r}) \bar{T}^{a}_{rs}\tilde{u}^{c}_{s}+  
\overline{\tilde{d}^{c}}_{r}\bar{T}^{a}_{rs}(\partial_{m}\tilde{d}^{c}_{s})-
(\partial_{m}\overline{\tilde{d}^{c}}_{r}) \bar{T}^{a}_{rs}\tilde{d}^{c}_{s})g^{a m}, \nonumber \\
\end{eqnarray}
which is the same of the MSSM, see Eq.(\ref{sqsqgmssm}), and the Feynman rule is given by Eqs.(\ref{squarksflavor},\ref{regrafeynman3}), as 
expected.

\subsubsection{Squark-Squark-Gluon-Gluon Interaction }
This interaction is given by the last line of Eq.(\ref{quarkinteslr}), and it is given by
\begin{eqnarray}
{\cal L}_{\tilde{q}\tilde{q}gg}&=&-g^{2}_{s}(
\bar{\tilde{u}}_{r}T^{a}_{rs}T^{b}_{st}\tilde{u}_{t}+
\bar{\tilde{d}}_{r}T^{a}_{rs}T^{b}_{st}\tilde{d}_{t}+
\overline{\tilde{u}^{c}}_{r}\bar{T}^{a}_{rs}\bar{T}^{b}_{st}\tilde{u}^{c}_{t}+
\overline{\tilde{d}^{c}}_{r}\bar{T}^{a}_{rs}\bar{T}^{b}_{st}\tilde{d}^{c}_{t})g^{a m}g^{b}_{m}, \nonumber \\
\end{eqnarray}
which is the same as Eq.(\ref{sgsgggmssm}), and the Feynman rule is given at Eq.(\ref{regrafeynman4}).

\subsubsection{Gluino-Quark-Squark Interaction}

This interaction is given by the third line of Eq.(\ref{quarkinteslr}), and can be rewritten as
\begin{eqnarray}
{\cal L}_{\tilde{q}qg}&=&- \sqrt{2}\imath g_{s}( \bar{u}_{r}T^{a}_{rs}\tilde{u}_{s}\overline{\lambda^{a}}_{C}-
\overline{\tilde{u}}_{r}T^{a}_{rs}u_{s}\lambda^{a}_{C}+  \bar{d}_{r}T^{a}_{rs}\tilde{d}_{s}\overline{\lambda^{a}}_{C}-
\overline{\tilde{d}}_{r}T^{a}_{rs}d_{s}\lambda^{a}_{C} \nonumber \\ &+&
 \overline{u^{c}}_{r}\bar{T}^{a}_{rs}\tilde{u}^{c}_{s}\overline{\lambda^{a}}_{C}-
\overline{\tilde{u}^{c}}_{r}\bar{T}^{a}_{rs}u^{c}_{s}\lambda^{a}_{C}+
\overline{d^{c}}_{r}\bar{T}^{a}_{rs}\tilde{d}^{c}_{s}\overline{\lambda^{a}}_{C}-
\overline{\tilde{d}^{c}}_{r}\bar{T}^{a}_{rs}d^{c}_{s}\lambda^{a}_{C}),
\end{eqnarray}
which is given by Eq.(\ref{sgqsq}), and the Feynman rule is given by Eq.(\ref{regrafeynman5}).

As a conclusion of sections 2 and 3, we have shown that the Feynman rules of the strong sector are the 
same in the following MSSM, NMSSM, MSSM3RHN and SUSYLR models.

\section{Parameters}
\label{spspoints}

The ``Snowmass Points and Slopes'' (SPS) \cite{sps1} are a set of benchmark points
and parameter lines in the MSSM parameter space corresponding to different scenarios in the search
for supersymmetry at present and future experiments (See \cite{{sps2}} for a very nice review). 
The aim of this convention
is reconstructing the fundamental supersymmetric theory, and its breaking
mechanism, from the experimental data. 

The different scenarious correspond to three different kinds of models. 
The points SPS 1-6 are Minimal Supergravity (mSUGRA) model, 
SPS 7-8 are gauge-mediated symmetry breaking (GMSB) model, and SPS 9 are 
anomaly-mediated symmetry breaking (mAMSB) model (\cite{sps1,sps2,sps}), see appendix \ref{apend:sps}. 

Each set of parameters leads to different masses of the 
gluinos and squarks, wich are the only relevant parameters in our study, 
and we shown their values in Tab.(\ref{tab:tmasses}). In this paper, all these ten possibilities will be considered in 
our predictions for gluino production.

\begin{table}[htb]
\renewcommand{\arraystretch}{1.10}
\begin{center}
\normalsize
 \vspace{0.5cm}
\begin{tabular}{|c|c|c|}
\hline
\hline
Scenario & $m_{\tilde{g}}\, (GeV)$ & $m_{\tilde{q}}\, (GeV)$ \\
\hline
\hline
 SPS1a & 595.2  & 539.9 \\
 SPS1b & 916.1  & 836.2 \\
 SPS2 & 784.4  & 1533.6 \\
 SPS3 & 914.3  & 818.3 \\
 SPS4 & 721.0  & 732.2 \\
 SPS5 & 710.3  & 643.9 \\
 SPS6 & 708.5  & 641.3 \\
 SPS7 & 926.0  & 861.3 \\
 SPS8 & 820.5  & 1081.6 \\
 SPS9 & 1275.2  & 1219.2 \\
  \hline
\hline
\end{tabular}
\caption{The values of the masses of gluinos and squarks in the SPS scenarios.}
\label{tab:tmasses} 
\end{center}
\end{table}

\section{Gluino Production}
\label{gluinoprod}

Gluino and squark production at hadron colliders occurs dominantly via strong interactions. Thus, their production rate 
may be expected to be considerably larger than for sparticles with just electroweak interactions whose 
production was widely studied in the literature \cite{dress,tata}. 
As shown above, the Feynman rules of the 
strong sector are the same in the followings MSSM, NMSSM, MSSM3RHN and SUSYLR
 models. Therefore the diagrams that contribute to the gluino production 
are the same in these models. In this sense, regarding the supersymmetric extensions of SM 
we consider here, the analysis we are going to do is model independent.

In the present paper we study the gluino production in $pp$
collisions. We will study the following reactions 
\begin{equation}
pp \longrightarrow \tilde{g}\tilde{g}, \tilde{g}\tilde{q} +X\,\,\,,
\end{equation}
where X is anything, in the proton--proton collisions at the LHC.

In order to make a consistent comparison and for sake of simplicity, we restrict ourselves to leading-order
(LO) accuracy, 
where the partonic cross-sections for the production of squarks and gluinos in hadron
collisions were calculated at the Born level already quite some time ago \cite{Dawson}. 
The corresponding NLO calculation has already been done for the MSSM case \cite{Zerwas}, and the
impact of the higher order terms is mainly on the normalization of the cross
section, which could be
taken in to account here by introducing a K factor in the results here obtained \cite{Zerwas}.

\begin{figure}[tb]
\begin{center}
\epsfig{file=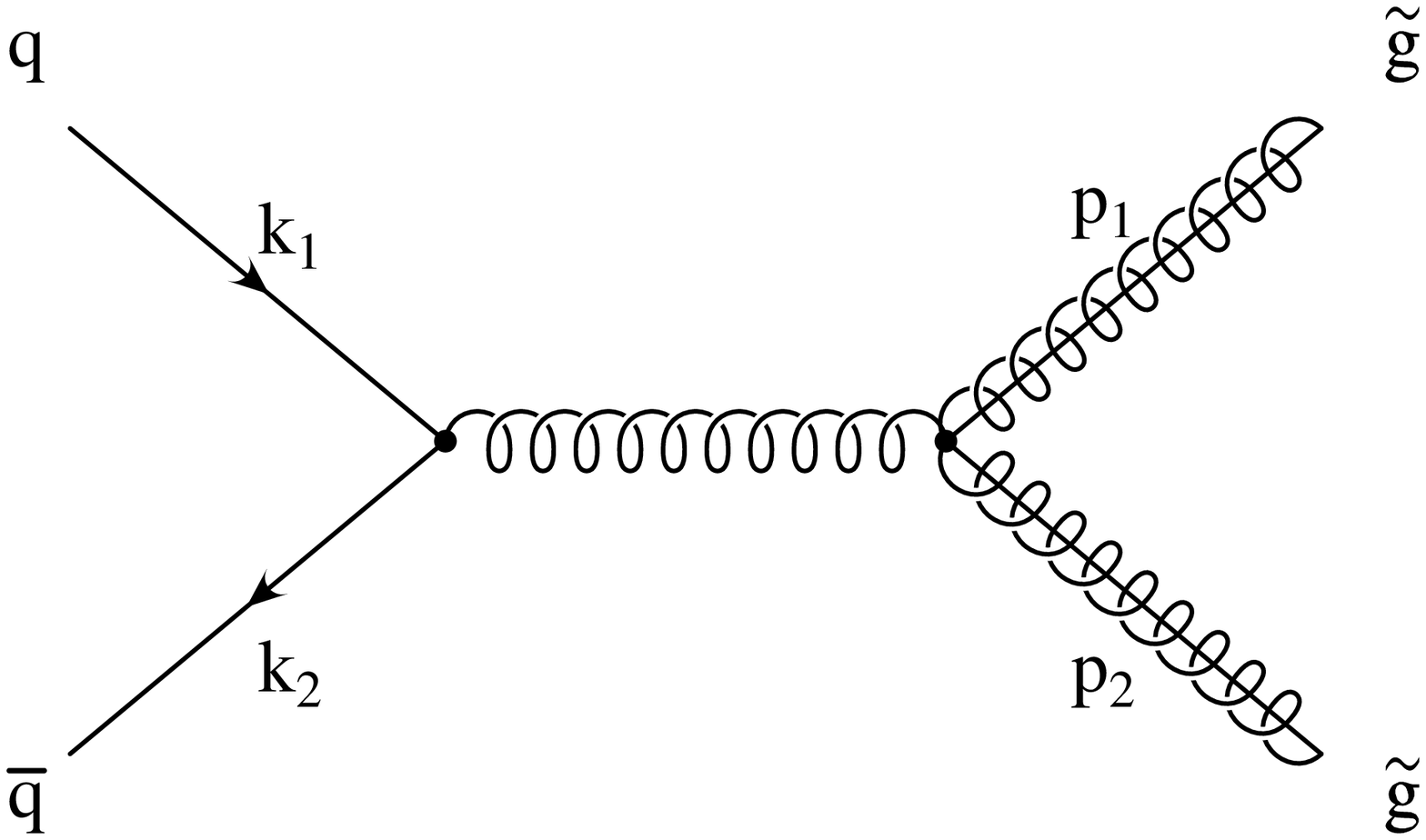,width=4cm}
\epsfig{file=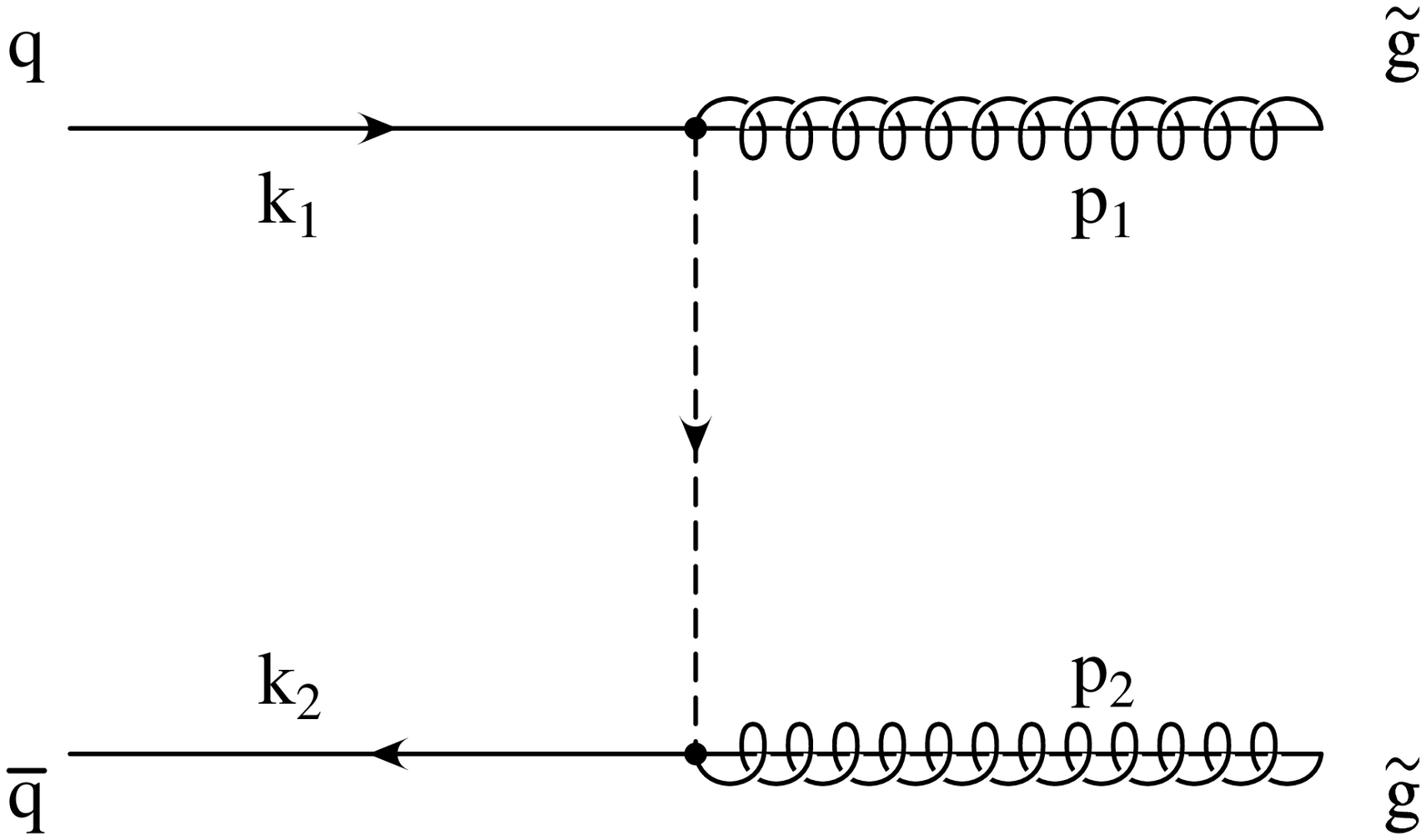,width=4cm}
\epsfig{file=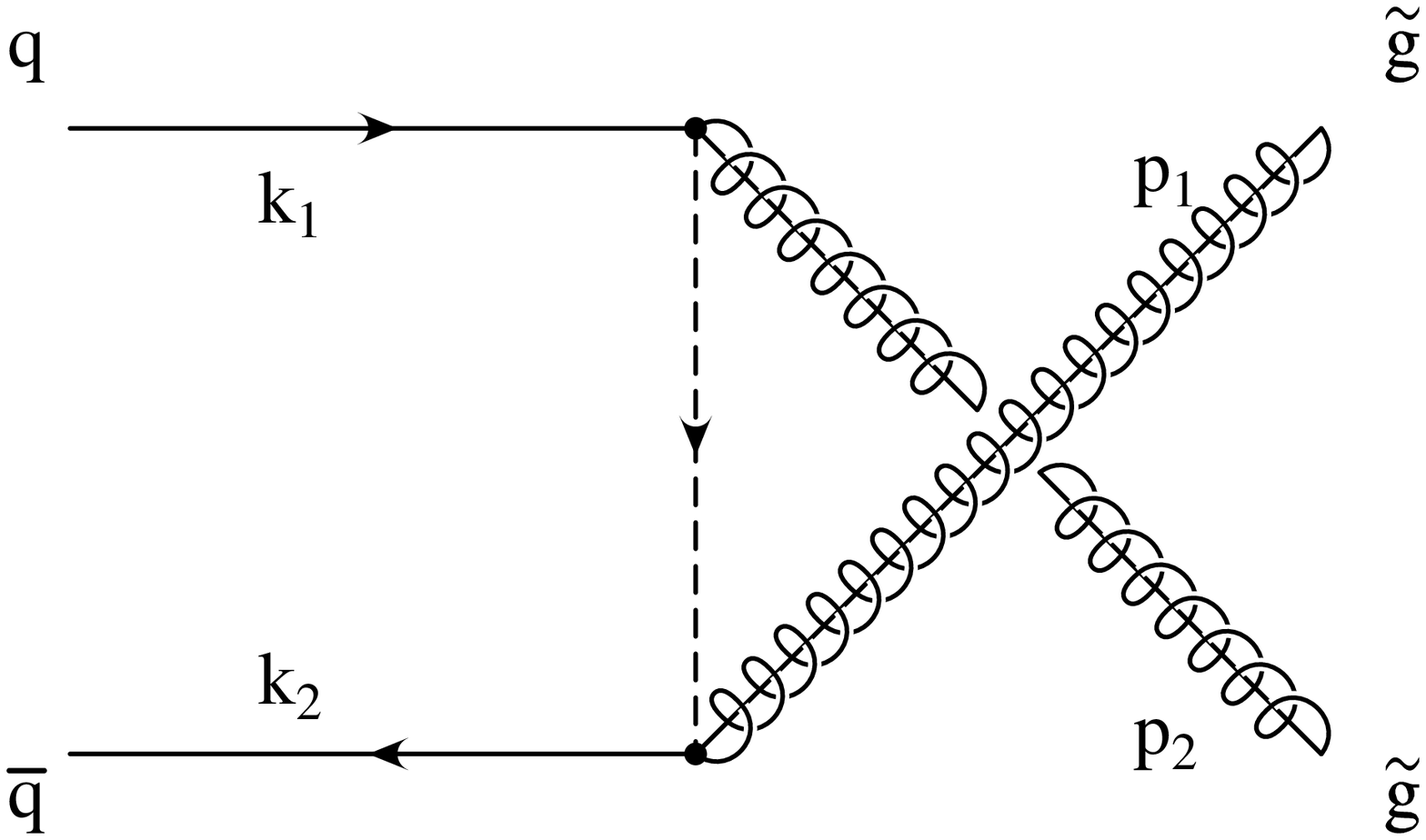,width=4cm}\\(a)\\
\epsfig{file=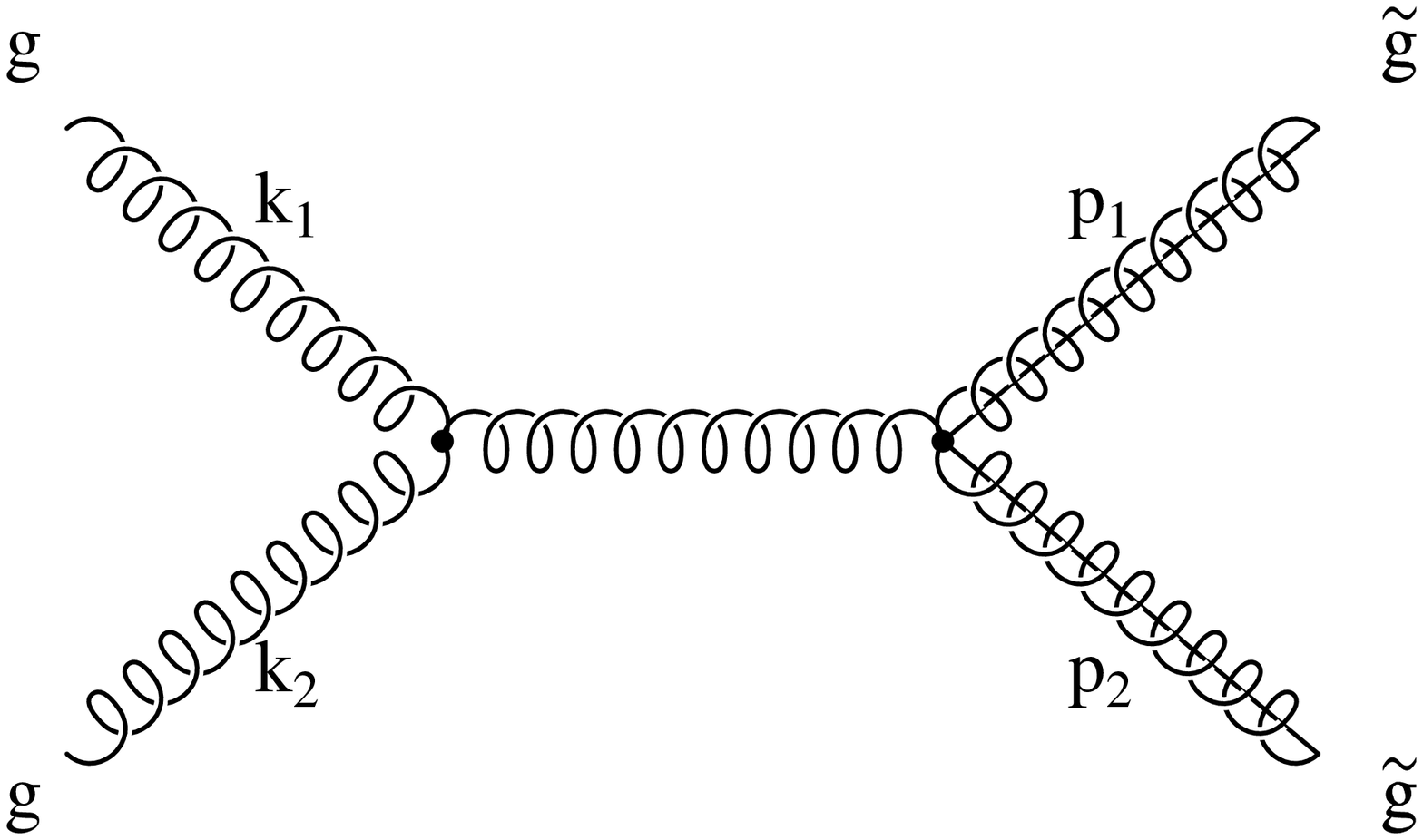,width=4cm}
\epsfig{file=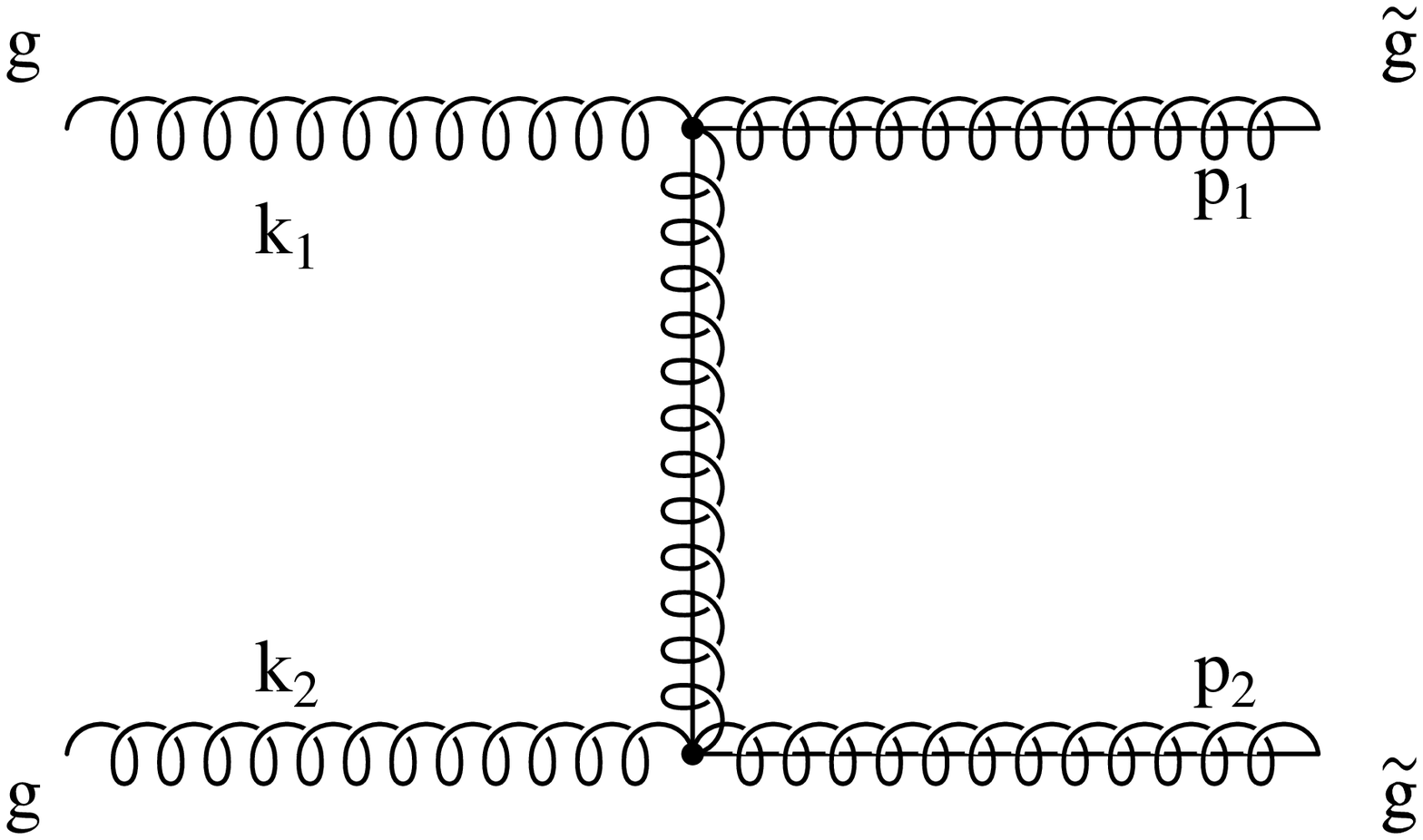,width=4cm}
\epsfig{file=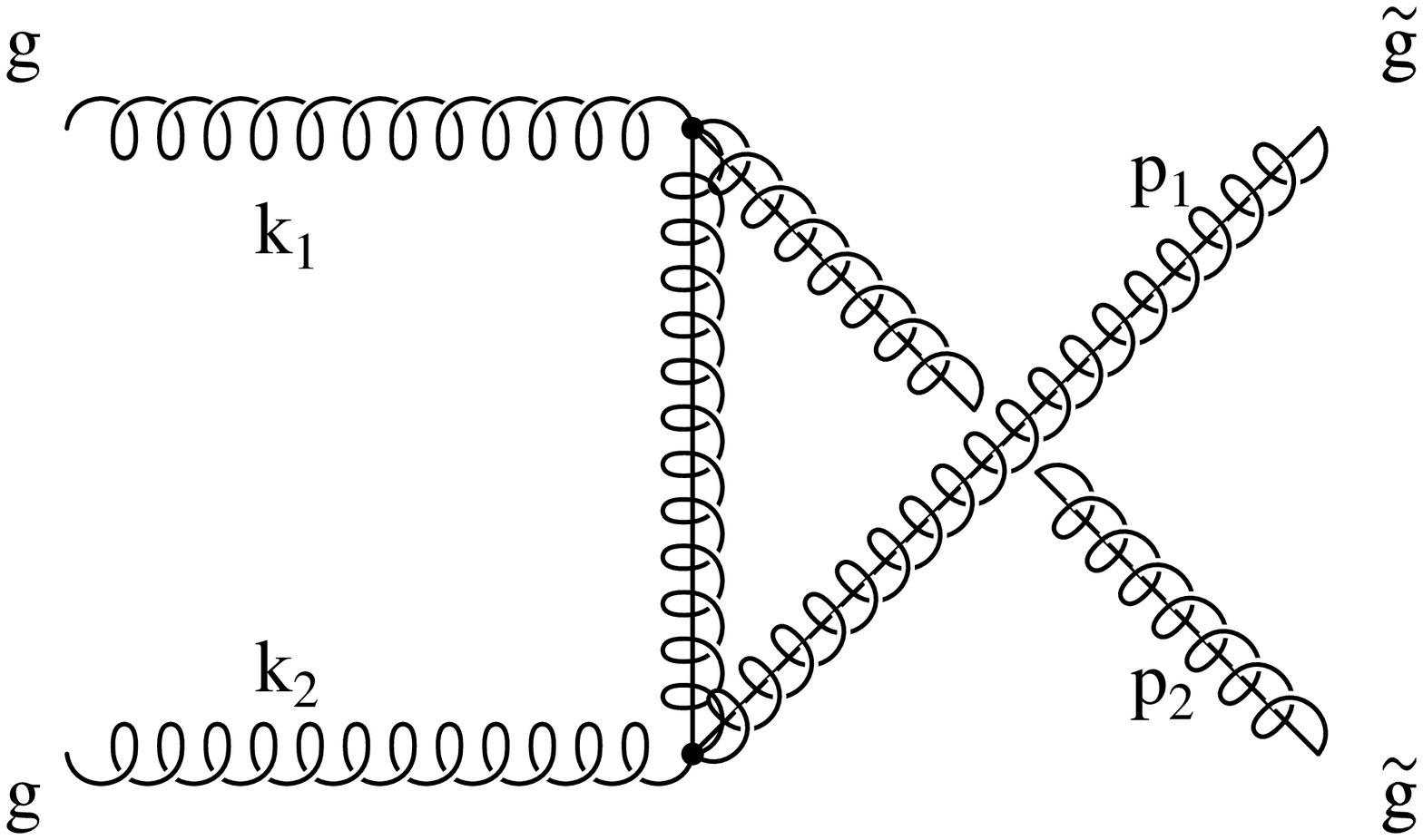,width=4cm}
\\(b)\\
\end{center}
\caption{Feynman diagrams for gluino pair production: (a)
  quark-antiquark initial states, (b) gluon-gluon initial states.}
\label{fig:GG}
\end{figure}

\begin{figure}[tb]
\begin{center}
\epsfig{file=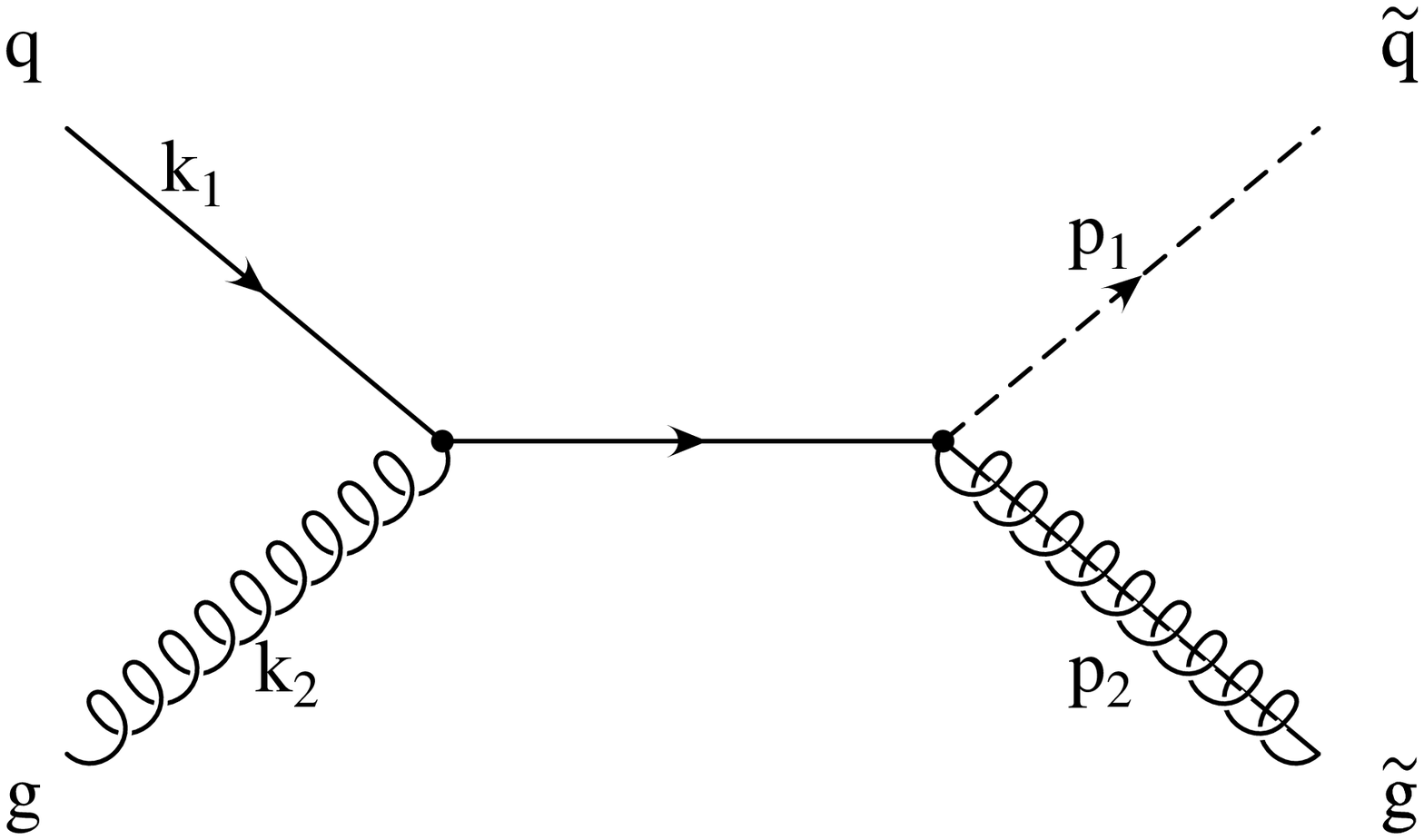,width=4cm}
\epsfig{file=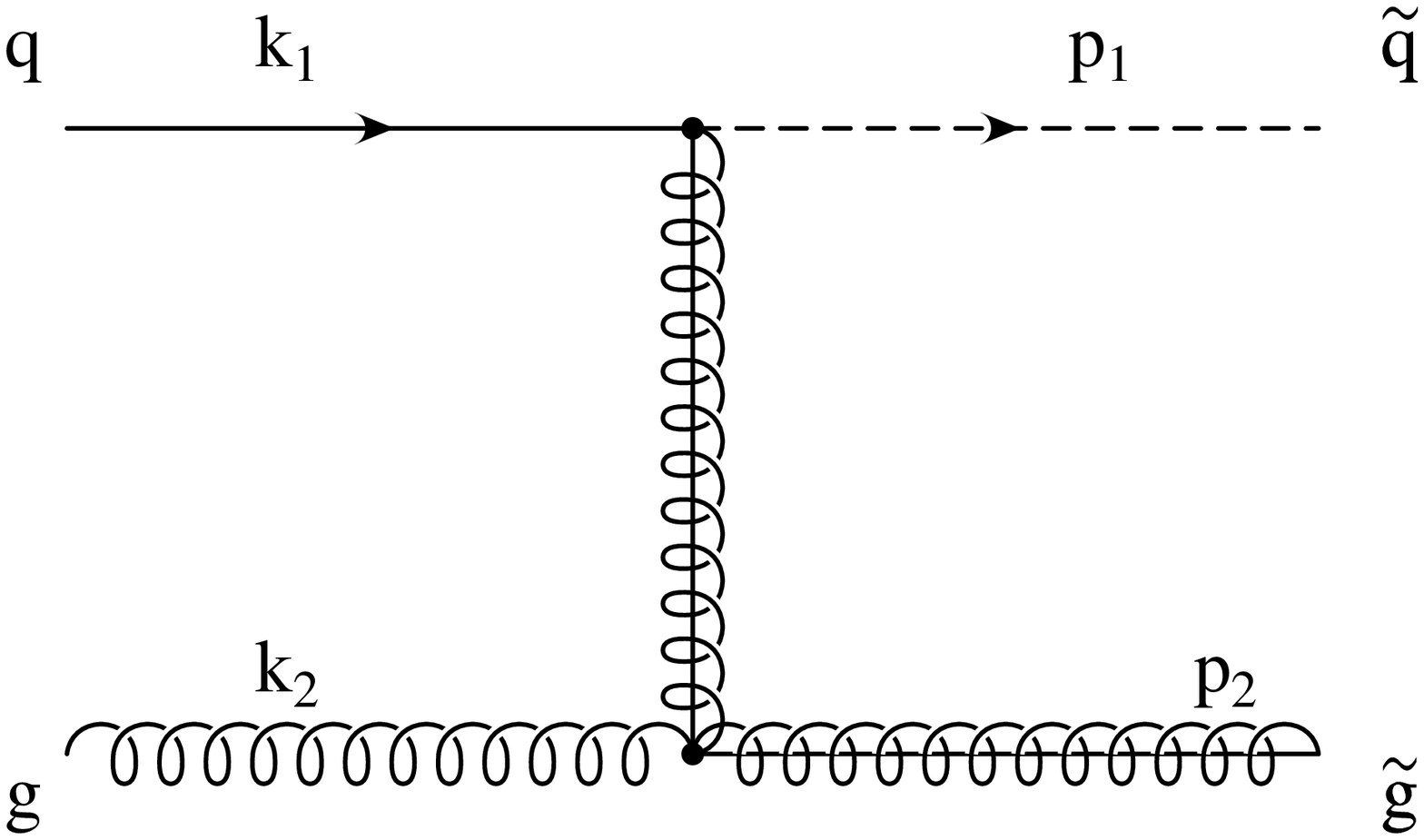,width=4cm}
\epsfig{file=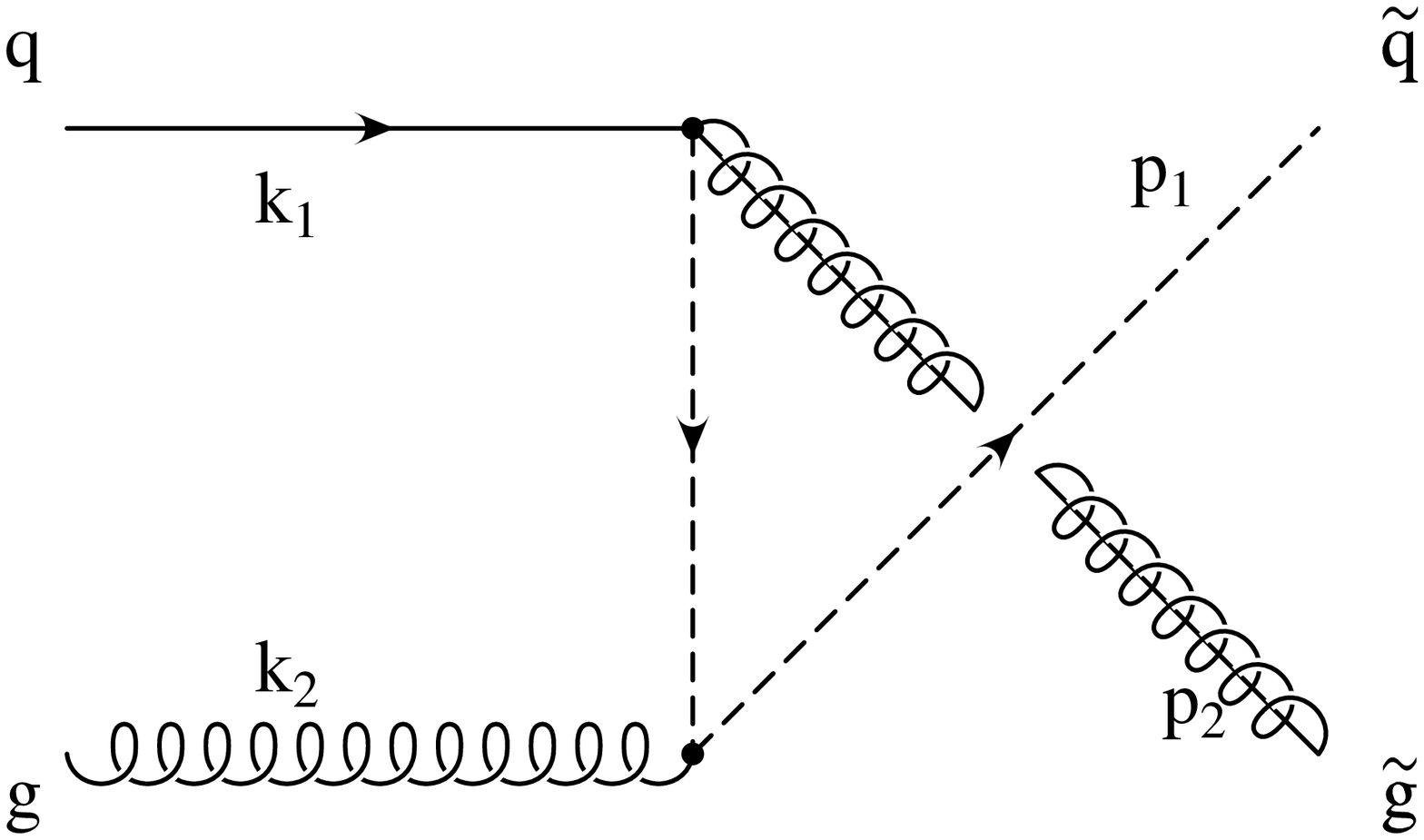,width=4cm}
\end{center}
\caption{Feynman diagrams for squark--gluino production.}
\label{fig:QG}
\end{figure}

The LO QCD subprocesses for single gluino production are gluon-gluon and quark-antiquark
anihilation ($g g  \rightarrow \tilde{g} \tilde{g}$ and $q \bar q  \rightarrow \tilde{g} \tilde{g}$) (shown in Fig.~\ref{fig:GG}),  
and the Compton process $qg\rightarrow \tilde{g}\tilde{q}$ (shown in Fig. ~\ref{fig:QG}).
For double gluino production only the anihilation processes contribute. 
These two kinds of events could be separated, in principle, by analysing the
different decay channels for gluinos and squarks \cite{dress,tata}.

Incoming quarks (including incoming $b$ quarks) are assumed to be massless,
such that we have $n_f=5$ light flavours. We only consider final state 
squarks corresponding to the light quark flavours. All 
squark masses are taken equal to $m_{\tilde q}$ 
\footnote{$L$-squarks and $R$-squarks are therefore mass-degenerate  
and experimentally indistinguishable.}. We do not consider in detail top squark
production where these assumptions do not hold and which require
a more dedicated treatment~\cite{plehn}.

The invariant cross section for single gluino production can be written as \cite{Dawson}
 \begin{eqnarray}
E\frac{d\sigma }{d^3p}= \sum_{ijd} \int_{x_{min}}^1 dx_a
f_i^{(a)}(x_a,\mu )f_j^{(b)}(x_b,\mu )\,\,\,\,\,\,\,\,\frac{x_ax_b}{x_a-x_{\perp} \left( \frac{\zeta + \cos \theta}{2\sin \theta} \right) }
\frac{d \hat{\sigma}}{d \hat{t}}(ij\rightarrow \tilde{g} d), 
\end{eqnarray}
where $f_{i,j}$ are the parton distributions of the incoming protons 
and $\frac{d \hat{\sigma}}{d \hat{t}}$ is the LO partonic cross section
\cite{Dawson} for the subprocesses involved. The identified gluino is produced at center-of-mass angle $\theta$ and 
transverse momentum $p_T$, and $x_{\perp}=\frac{2p_T}{\sqrt{s}}$. The Mandelstam variables  
of the partonic reactions $ij\rightarrow \tilde{g}\tilde{g}, \tilde{g}\tilde{q}$ 
are then
\begin{eqnarray}
\hat{s}&=& x_ax_bs, \nonumber \\ 
\hat{t}&=& m_{\tilde{g}}^2-x_ax_{\perp}s\left( \frac{\zeta-\cos \theta}{2\sin \theta}\right), 
\nonumber \\ 
\hat{u}&=& m_{\tilde{g}}^2-x_bx_{\perp}s\left(\frac{\zeta+\cos \theta}{2\sin \theta}\right).
\end{eqnarray}
Here
\begin{eqnarray}
x_b&=&\frac{2\upsilon + x_ax_{\perp}s\left( \frac{\zeta-\cos \theta}{\sin \theta}\right)
}{2x_as-x_{\perp}s\left(\frac{\zeta+\cos \theta}{\sin \theta}\right) },
\nonumber \\ 
x_{min}&=&\frac{2\upsilon + x_{\perp}s\left(\frac{\zeta+\cos \theta}{\sin \theta}\right) }
{2s-x_{\perp}s\left( \frac{\zeta-\cos \theta}{\sin \theta}\right) }, \nonumber \\
\zeta &=& {\left( 1+\frac{4m_{\tilde{g}}^2\sin ^2
\theta}{x_\perp^2s} \right)}^{1/2}, \nonumber \\ 
\upsilon &=& m_d^2-m_{\tilde{g}}^{2}, 
\end{eqnarray}
where $m_{\tilde{g}}$ and $m_d$ are the masses of the final-state partons
produced. 
The center-of-mass angle $\theta$ and the differential cross section above
can be easilly written in terms of the pseudorapidity variable $\eta=-\ln \tan
(\theta/2)$,
 which is one of the experimental observables\footnote{Remember that since gluinos are heavy, their rapidity and pseudorapidity are not equal.}.
The total cross section for the gluino production can be obtained from above
upon integration. The corresponding partonic total cross sections for the
subprocesses considered are well known and can be found at \cite{Dawson,Zerwas}.

\section{Numerical Results}
\label{results}

Here in this section we present our numerical results and plots about the 
gluino production at the LHC. Since the $pp$ CM energy $\sqrt{s}$ =14 TeV is
several times larger than the expected gluino and squark masses, these particles
might be produced and detected at the LHC.

\begin{figure}[tb]
\begin{center}
\epsfig{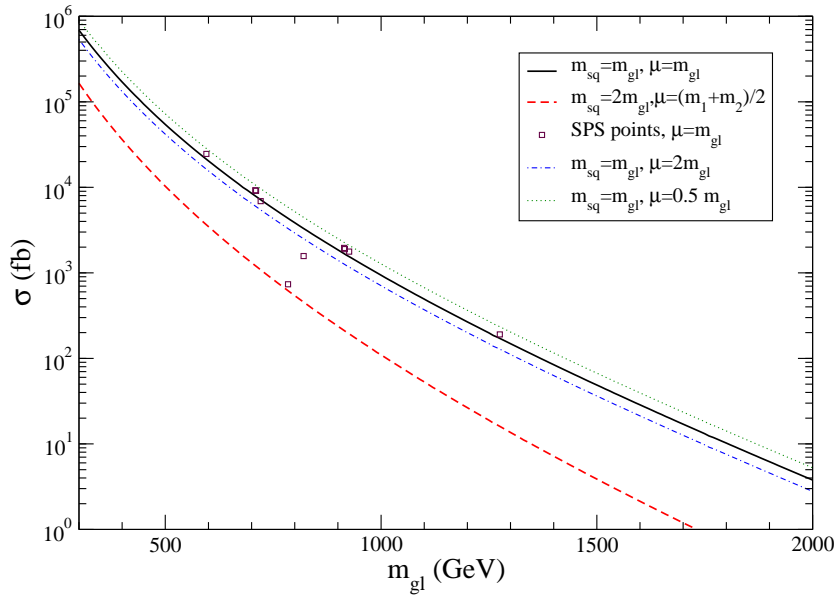}
\end{center}
\caption{The total LO cross section for gluino production at the LHC as a function of the gluino masses. Parton densities: CTEQ6L, with two assumptions on the squark masses and choices of the hard scale (curves). The sensitivity with the hard scale is also presented in the case $m_{\tilde{q}}=m_{\tilde{g}}$. The points are the numerical results for the SPS points as explained in the text.}
\label{fig:sigtot}
\end{figure}

In Fig.\ref{fig:sigtot} we present the LO QCD total cross section for gluino 
production at the LHC as a function of the gluino masses. 
We use the CTEQ6L \cite{Pumplin:2002vw}, parton densities, with two assumptions on the squark masses and 
choices of the hard scale (curves). The sensitivity with the hard scale is also presented in the case 
$m_{\tilde{q}}=m_{\tilde{g}}$.  We also, want to stress that the behaviour of our curves are similar 
to ones presented at Chapter 12 on reference \cite{tata}, where they use the CTEQ5L parton distribution on their calculation.

The search for gluinos and squarks (as well as other searches for SUSY particles) and the
possibility of detecting them will depend on their real masses. 
We also show (points) in Fig.\ref{fig:sigtot} the numerical results for the LO
gluino total cross section in all SPS scenarios, 
fixing the gluino and squark masses, taken from Tab.(\ref{tab:tmasses}).

The results show a strong dependence on the masses of gluinos and
squarks. In the model curves, we get a larger cross section in the degenerated
mass case, which agrees with \cite{tata}. Most of the SPS points are close to
the first curve, which can be easily understood by looking at Table \ref{tab:tmasses}.

\begin{figure}[tb]
\begin{center}
\epsfig{file=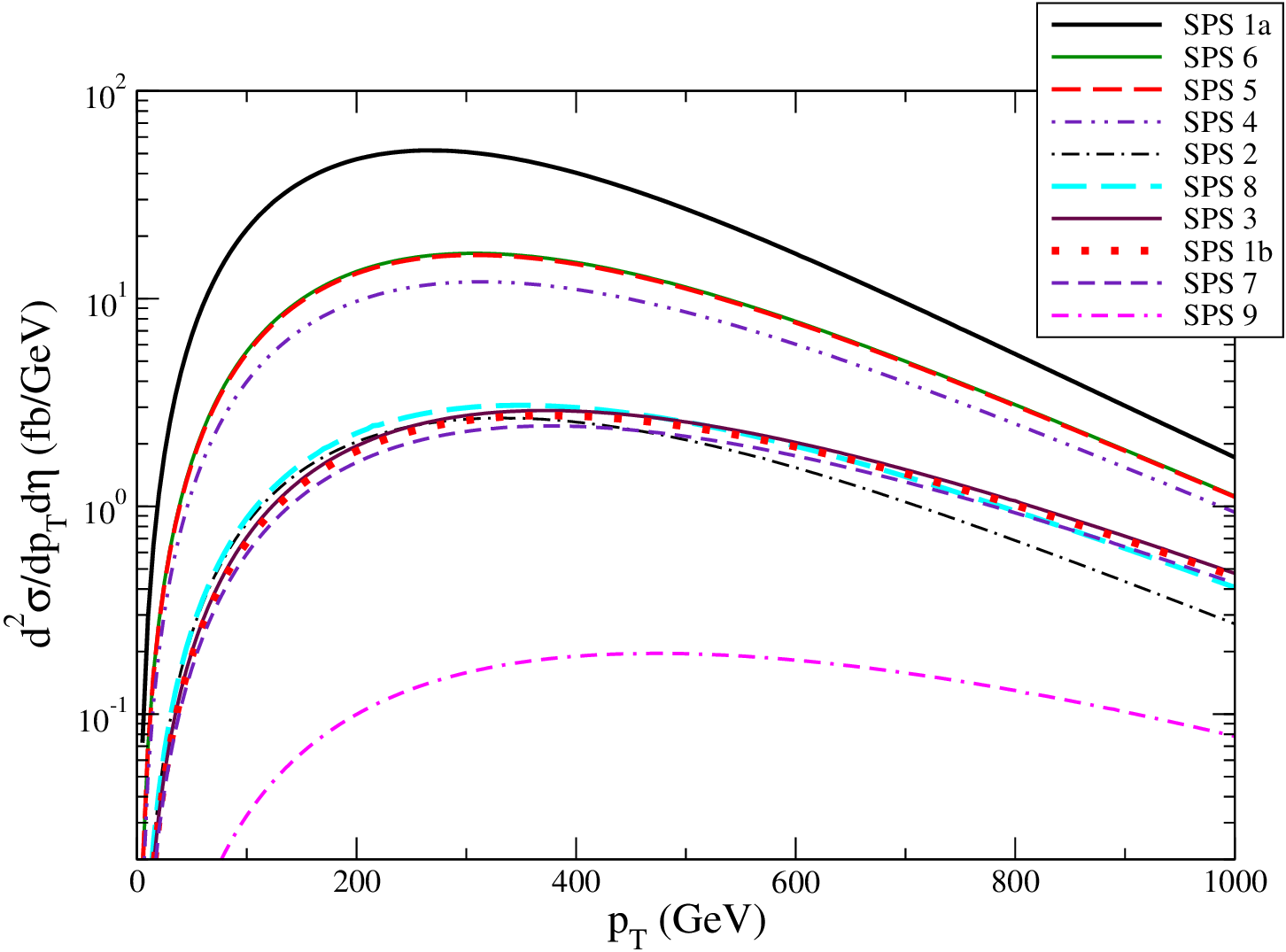,width=13cm}
\end{center}
\caption
{The LO $p_T$ distributions for single gluino production at the LHC ($|\eta |<2.5$) for the different SPS points  \cite{sps1,sps2}. We use CTEQ6L parton densities, and $\mu^2=m_{\tilde{g}}^2+p_T^2$ as a hard scale.}
\label{fig:sigpt}
\end{figure}

To discriminate among the different scenarious, it is relevant to look into 
more detailed observables such as differential distributions. 
In Fig.\ref{fig:sigpt} we present the transverse momentum distributions for single
gluino production at LHC energies. 
The results show a huge diference in the magnitude for different scenarios - 
 SPS1a (mSUGRA) gives the bigger values, SPS9 (AMSB) the smallest one. The predictions for the points SPS5 and 
 SPS6 (both mSUGRA) are indistinguishable, since the gluinos and squarks masses are almost the same in these two mSUGRA scenarios, so gluino production is not a good process to discriminate between them. Regarding the magnitude of the cross section of gluino production, this process could be usefull to discriminate among basically four kinds of SPS scenarios, namelly:\\ 
  (i) SPS 1a; \\
  (ii) SPS 5, SPS 6, and SPS 4 (in fact, this point is a bit lower than the other two);\\
  (iii) SPS 1b, SPS 2, SPS 3, SPS 7, SPS 8; \\
  (iv) SPS 9.
 
Looking into the details of the predictions, namely the behavior of the cross section, can give us further information to discriminate among the (iii)-scenarios. The $p_T$ dependency in these scenarios is not the same for most of these points (except SPS 1b and SPS 3, which are almost equal). The SPS 2 (mSUGRA) model prediction has clearly a steeper falloff at high very high $p_T$, and the reason for that is the much higher squark mass in this scenario. At moderate $p_T$ of about 200 GeV, the SPS 7 (GMSB) curve has the lowest normalization, that is because the gluino mass is higher in this scenario. However, because the squark mass is much lower than in SPS 2, the falloff at higher $p_T$ is less steep and in this region the curve tend to other (iii)-points. So, the gluino and squark masses are more important in different regions and there is an interplay between them wich produces different behaviors in different scenarios.

The gluino mass is important in all subprocesses, but the squark mass only contributes to the $q\bar{q}$ anihilation and the Compton-like process $qg\to \tilde{g}\tilde{q}$, because of the t-channel squark exchange in both subprocesses and, of course, the squark production in the latter one. Comparing these two processes, the Compton process is dominant. Therefore, the different behaviors of the cross secion are mainly due to the Compton-like contribution.

\begin{figure}[t]
\begin{center}
\epsfig{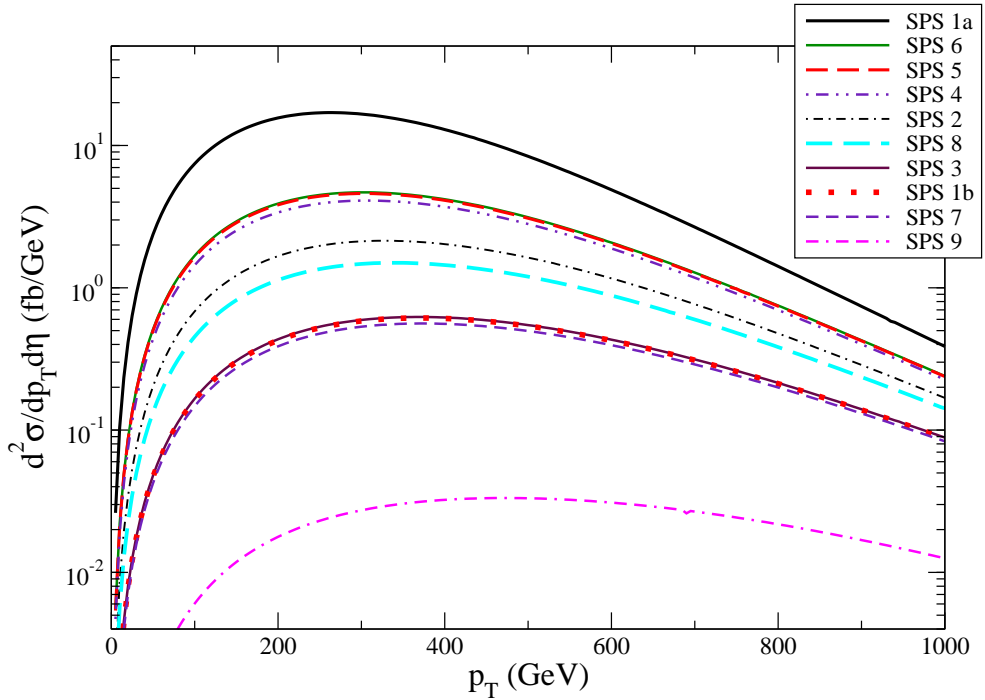}
\end{center}
\caption{The LO $p_T$ distributions for double gluino production at the LHC ($|\eta |<2.5$) for the 
different SPS points 
\cite{sps1,sps2}. We use CTEQ6L parton densities, and $\mu^2=m_{\tilde{g}}^2+p_T^2$ as a hard scale.} 
\label{fig:sigpt2g}
\end{figure}

A complementary analysis can be done by considering double gluino production, which is easy to obtain from the calculation above, by picking only the anihilation processes (Fig. \ref{fig:GG}). 
The results for double gluino production are shown in Fig. \ref{fig:sigpt2g}. 
Similarly to the previous case, the results show huge diferences in the magnitude of the cross section for different scenarios - 
 SPS1a gives the bigger values, SPS9 the smallest one. Also, we find very close values for SPS1b, SPS3 (mSUGRA) and SPS7 (GMSB), which makes it difficult to discriminate between these mSUGRA and GMSB models. 
 The same occurs for SPS5 and SPS6 (both mSUGRA). 
However, differently from the single gluino case, the $p_T$ dependencies are similar in all scenarios. Another difference is the magnitude of curves for the points SPS 2 and SPS 8, wich can be clearly separated from the other (iii)-points described above. From all this, we conclude that both processes, single and double gluino production, are complementary and usefull to make different discriminations among the SPS scenarios.

\section{\textbf{Conclusions}}

\label{sec:concl}

In this paper we have studied the color sector in some extensions of the SM, namely MSSM and SUSYLR, and some others. We have derived the Feynman rules for the strong sector, and have showed explicitly that they are the same in all these models. This happens because the strong sector is the same in all SUSY extensions considered. 

There are several scenarios for SUSY breaking, within the SPS convention, which imply in different values for the masses of the supersymmetric particles. To find the correct SUSY breaking mechanism one has to consider different observables which could be measured 
when SUSY particles starts to be detected. In this article, we analyse gluino production at the LHC.  

Gluinos are color octet fermions and play a major role to understanding sQCD. 
Because of their large mass as predicted in several scenarios, up to now the LHC is the only possible machine where they could be found.
 
Because the Feynman rules for the strong sector are the same in all SM extensions considered, the gluino production cross section are indeed equal in these models. Our results are in this sense model independent, or conversely, gluino production is not a good process to discriminate among those SUSY extensions of SM.

Besides, our results depend on the gluino and squark masses and no other SUSY parameters. Since the masses of gluinos come only from the soft terms, measuring their masses can test the soft SUSY breaking approximations. We have considered all the SPS scenarios and showed the corresponding differences on the magnitude of the production cross sections. 
From this it is easy to distinguish 
mAMSB from the other scenarios. However, it is not so easy to distinguish mSUGRA from GMSB depending on the real values of masses of gluinos and squarks. Gluino production cannot distinguish the two scenarios SPS1b and SPS7, provided the gluino and squark masses are almost similar in these two cases (the same occurs for SPS 5 and SPS 6). For the other scenarios, such discrimination
can be done, especially if we consider both single and double gluino production as complementary processes.

Gluino production is not a good process to discriminate among the Supersymmetric Models, but can be helpfull is determining the correct SUSY breaking scenario and to understanding supersymmetric quantum chromodynamics.

\begin{center}
{\bf Acknowledgments} 
\end{center}
This work was partially financed by the Brazilian funding agency
CNPq, CBM under contract number 472850/2006-7, and MCR under contract number 
309564/2006-9. We would like to thank V. P. Gon\c calves for useful discussion.  We are gratefull 
to Pierre Fayet for many useful discussions and also for providing us with very useful material about 
the early days of phenomenology of Supersymmetric QCD.

\appendix

\section{Tables of SPS Convention}
\label{apend:sps}

On this appendix we present the parameters of the SPS convention, which are given at Table \ref{tab:params}.

\begin{table}[h]
\begin{center}
\begin{tabular}{|ccccccc|}
\hline\hline
SPS & \multicolumn{6}{c}{Point \hspace{3em}} \\ 
\hline\hline
mSUGRA: & $m_0$ & $m_{1/2}$ & $A_0$ & $\tan\beta$ & & \\
\hline
1a & 100  & 250 &   -100 & 10 & &   \\
1b & 200  & 400 &      0 & 30 & &  \\
2  & 1450 & 300 &      0 & 10 & & \\
3  &   90 & 400 &      0 & 10 & & \\
4  &  400 & 300 &      0 & 50 & &  \\
5  &  150 & 300 &  -1000 &  5 & &  \\
\hline\hline
mSUGRA-like: & $m_0$ & $m_{1/2}$ & $A_0$ & $\tan\beta$ & 
               $M_1$ & $M_2 = M_3$  \\
\hline      
6  &  150 & 300 &      0 & 10 & 480 & 300  \\ 
\hline\hline         
GMSB: & $\Lambda/10^3$ & $M_{\rm mes}/10^3$ & $N_{\rm mes}$ & $\tan\beta$ & &  \\ 
\hline     
7  &   40 &  80 &      3 & 15 & &  \\    
8  &  100 & 200 &      1 & 15 & &  \\   
\hline\hline 
AMSB: & $m_0$ & $m_{\rm aux}/10^3$ & & $\tan\beta$ & & \\ 
\hline         
9  &  450 &  60 &        & 10 & & \\
\hline\hline   
\end{tabular}
\caption{The parameters for the Snowmass Points and Slopes (SPS). On this table all the scenarios consider $\mbox{sign}\mu =+$ taken from \cite{sps2}.}
\label{tab:params}
\end{center}\nonumber

\end{table}



\begin{thebibliography}{99} 

\bibitem{sg}S. L. Glashow, {\sl Nucl. Phys.}{\bf 22}, 579 (1961);
S. Weinberg, {\sl Phys. Rev. Lett.}{\bf 19}, 1264 (1967);
A. Salam in {\sl Elementary Particle Theory: Relativistic Groups
and Analyticity}, Nobel Symposium N8 (Alquivist and Wilksells, Stockolm,
1968);
S. L. Glashow, J.Iliopoulos and L.Maini,  {\sl Phys. Rev.}{\bf D 2}, 1285 (1970).

\bibitem{super}
Y. A. Golfand and E. P. Likhtman, {\sl JETP Letters} {\bf13}, 452 (1971);
D. V. Volkov and V. P. Akulov,  {\sl JETP Letters} {\bf 16}, 621 (1972);
J. Wess and B. Zumino, {\sl Phys. Lett.} {\bf B49}, 52 (1974).

\bibitem{ssm} P. Fayet,  {\sl Phys. Lett.} {\bf B64} 159 (1976);
{\bf B69} 489 (1977).

\bibitem{susy}
M. F. Sohnius, {\sl Phys. Rep.} {\bf 128}, 41 (1985);
H. P. Nilles, {\sl Phys. Rep.} {\bf 110},1 (1984);
A. B. Lahanas and D. V. Nanopoulos, {\sl Phys. Rep.} {\bf 145}, 1 (1987);
S. J. Gates, M. Grisaru, M. Ro\v{c}ek and W. Siegel, {\it Superspace or One Thousand
and One Lessons in Supersymmetry} (Benjamin \& Cummings, 1983);\\
P.West, {\it Introduction to supersymmetry and supergravity}, 
2nd edition, World Scientific Publishing Co. Pte. Ltd., 
Singapore, (1990);\\
S.Weinberg, {\it The quantum theory of fields. Vol. 3. Supersymmetry} 
1st edition, Cambridge University Press, United Kindom, (2000).

\bibitem{wb}J. Wess and J. Bagger, {\it Supersymmetry and Supergravity}
2nd edition, Princeton University Press, Princeton NJ, (1992).

\bibitem{Fayet:2001xk}
  P.~Fayet,
{\sl Nucl. Phys. Proc. Suppl.}  {\bf 101},  81 (2001).


\bibitem{Chung:2003fi} D.~J.~H.~Chung, L.~L.~Everett, G.~L.~Kane,
S.~F.~King, J.~D.~Lykken and L.~T.~Wang, 
{\sl Phys.Rept.}{\bf 407}, 1 (2005). 


\bibitem {dress}M. Dress, R. M. Godbole and P. Royr, \textit{Theory and
Phenomenology of Sparticles} 1st edition, World Scientific Publishing Co. Pte.
Ltd., Singapore, (2004).

\bibitem {tata}H. E. Baer and X. Tata, \textit{Weak Scale Supersymmetry} 1st
edition, Cambridge University Press, United Kindom, (2006).

\bibitem{sugra} P. Nath and R. Arnowitt, {\em Phys. Lett.} {\bf B56}, 177 (1975);
D. Z. Freedman, P. van Nieuwenhuizen and S. Ferrara, {\em Phys. Rev.} {\bf D13}, 3214 (1976); 
S. Deser and B. Zumino, {\em Phys. Lett.} {\bf B62}, 335 (1976); see also
"Supersymmetry", S.Ferrara, ed. (North Holland/World Scientific, Amsterdam/Singapore,
1987).

\bibitem{ABF}
U. Amaldi, W. de Boer, H. F\"{u}rstenau, Phys. Lett. \textbf{B260}, 447 (1991).


\bibitem{INO82a}
   K. Inoue, A. Komatsu and S. Takeshita,
        {\sl Prog. Theor. Phys.} {\bf 68}, 927 (1982); 
        {\sl Prog. Theor. Phys.} {\bf 70}, 330 (1983).

\bibitem{running}
 V. Barger, M. S. Berger and P. Ohmann, {\em Phys. Rev.} {\bf D47}, 
1093 (1993); 
 W. de Boer, R. Ehret and D. Kazakov, {\em Z. Phys.}
 {\bf C67},  647 (1995);
W. de Boer et al., {\em Z. Phys.} {\bf C71}, 415 (1996).

\bibitem{Ibanez:wd}
  L.~E.~Iba\~nez and G.~G.~Ross,
  {\sl Phys. Lett.} {\bf B131}, 335 (1983);
B.~Pendleton and G.~G.~Ross,
  {\sl Phys. Lett.}  {\bf B98}, 291 (1981).

\bibitem{Dimopoulos:1981yj}
  S.~Dimopoulos, S.~Raby and F.~Wilczek,
  {\sl Phys. Rev.}  {\bf D24},  1681 (1981);
S.~Dimopoulos and H.~Georgi,
  {\sl Nucl. Phys.}  {\bf B193}, 150 (1981);
L.~E.~Iba\~nez and G.~G.~Ross,
  {\sl Phys. Lett.}  {\bf B105}, 439 (1981);
M.~B.~Einhorn and D.~R.~T.~Jones,
  {\sl Nucl. Phys.}  {\bf B196}, 475 (1982).

\bibitem{Kane:1992kq}
  G.~L.~Kane, C.~F.~Kolda and J.~D.~Wells,
  {\sl Phys. Rev. Lett.}  {\bf 70}, 2686 (1993);
 J.~R.~Espinosa and M.~Quiros,
  {\sl Phys. Lett.} {\bf B302}, 51 (1993).

\bibitem{lepewwg}   
LEP Electroweak Working Group, LEPEWWG/2001-01.

\bibitem{grav} P. Fayet, {\sl Phys. Lett.} {\bf B70}, 461 (1977).

\bibitem {mssm}H. E. Haber and G. L. Kane, {\sl Phys. Rep.}{\bf 117}, 75 (1985).

\bibitem{R} 
P. Fayet, {\sl Nucl. Phys.}  {\bf B90}, 104 (1975).


\bibitem{rp} P. Fayet, in New Frontiers in High-Energy Physics,
Proc. Orbis Scientiae, Coral Gables (Florida, USA), 1978,
eds. A. Perlmutter and L.F. Scott (Plenum, N.Y., 1978) p. 413.

\bibitem{ff} 
G.R. Farrar and P. Fayet,  {\sl Phys. Lett.} {\bf B76}, 575 (1978).

\bibitem{ff2} 
G.R. Farrar and P. Fayet, {\sl Phys. Lett.} {\bf B79}, 442 (1978);
{\sl Phys. Lett.} {\bf B89}, 191 (1980).


\bibitem{hall0}
L.~Hall and M.~Suzuki,
{\em Nucl. Phys.}, {\bf B231}, 419,(1984);
H.P. Nilles and N. Polonsky, {\sl Nucl. Phys.} {\bf B499}, 33 (1997);
 T. Banks, Y. Grossman, E. Nardi and Y. Nir,  
{\sl Phys. Rev.} {\bf D52}, 5319 (1995);
 F.M. Borzumati, Y. Grossman, E. Nardi and Y. Nir,  
 {\sl Phys.Lett.} {\bf B384}, 123 (1996);
 E.Nardi, {\sl Phys. Rev.} {\bf D55}, 5772 (1997);
Y. Grossman and H. E. Haber, {\it Phys. Rev. Lett.} 
 {\bf 78}, 3438 (1997);
{\it   Phys.Rev.} {\bf D59}, 093008;  hep-ph/9906310.

\bibitem{dreiner}H. Dreiner, hep-ph/9707435;
G. Bhattacharyya, {\sl Nucl. Phys. Proc. Suppl.} {\bf 52A}, 83 (1997); 
hep-ph/9709395; B. Allanach, A. Dedes and H. Dreiner, 
{\it Phys. Rev.} {\bf D60}, 075014 (1999).

\bibitem{barbier} R. Barbier {\it et al.}, Phys. Rept. {\bf 420},
1 (2005).
G. Moreau, [arXiv:hep-ph/0012156].

\bibitem{Fayet:1974pd}
  P.~Fayet,
  {\sl Nucl. Phys.}   {\bf B90}, 104 (1975).

\bibitem{kim} J.E. Kim and H.P. Nilles, {\sl Phys. Lett.} {\bf B 138}, 150 (1984).

\bibitem{minkowski} P. Minkowski, {\sl Phys.Lett.}{\bf B67}, 421 (1977).

\bibitem{grs79}
M.~Gell-Mann, P.~Ramond and R.~ Slansky, in Supergravity, edited by P.~ van Niewenhuizen and 
D.~Freedman, North Holland, Amsterdam,1979.

\bibitem{ms80a}
R.~N.~Mohapatra and G.~Senjanovic,
  {\sl Phys. Rev. Lett.} {\bf 44}, 912 (1980).

\bibitem{king}
S. F. King, hep-ph/9806440;
S.~Davidson and S.~F.~King,
  {\sl Phys. Lett.} {\bf B445}, 191 (1998);
S.~Antusch, E.~Arganda, M.~J.~Herrero and A.~M.~Teixeira,
  {\sl Nucl. Phys. Proc.Suppl.} {\bf 169}, 155 (2007).
  

\bibitem{Teixeira:2007gq}
  A.~M.~Teixeira, S.~Antusch, E.~Arganda and M.~J.~Herrero,
{\it In the Proceedings of 5th Flavor Physics and CP Violation Conference (FPCP 2007), Bled, Slovenia, 12-16 May 2007, pp 029}
  [arXiv:0708.2617 [hep-ph]].

\bibitem{ps74} J.C. Pati and A. Salam, {\sl Phys. Rev.} {\bf D10}, 275 
             (1974); 
               R.N. Mohapatra and J.C. Pati, {\it ibid} {\bf D11},
               566; 2558 (1975); G. Senjanovi\'{c} and R.N.
              Mohapatra, {\it ibid} {\bf D12}, 1502 (1975).
              

\bibitem{Huitu:1999qx}
  K.~Huitu, J.~Maalampi, P.~N.~Pandita, K.~Puolamaki, M.~Raidal and N.~Romanenko,
  arXiv:hep-ph/9912405.

\bibitem{melfo1} A. Melfo and G. Senjanovi\'c,  {\sl Phys.Rev.}{\bf D68}, 035013 (2003).

\bibitem{s79}
  G.~Senjanovic,
  {\sl Nucl. Phys.}  {\bf B153}, 334 (1979).


\bibitem{p74}
A.~M.~Polyakov,
  {\sl JETP Lett.}  {\bf 20}, 194 (1974)
  [Pisma Zh.\ Eksp.\ Teor.\ Fiz.\  {\bf 20} (1974) 430];
G.~'t Hooft,
  {Nucl. Phys.}  {\bf B79}, 276 (1974);
A.~Sen,
  {\sl Phys. Lett.} {\bf B153}, 55 (1985).

\bibitem{pss83}
J.~C.~Pati, A.~Salam and U.~Sarkar,
  {\sl Phys. Lett.}  B{\bf B133}, 330 (1983) .

\bibitem{mm80}
R.~N.~Mohapatra and R.~E.~Marshak,
  {\sl Phys. Rev. Lett.}  {\bf 44}, 1316 (1980)
  [Erratum-ibid.  {\bf 44}, 1643 (1980)];
Z.~Chacko and R.~N.~Mohapatra,
  {\sl Phys. Rev.}  {\bf D59}, 055004 (1999).

\bibitem{fy86}
M.~Fukugita and T.~Yanagida,
  {\sl Phys. Lett.}  {\bf B174}, 45 (1986).


\bibitem {susylr}K. Huitu, J. Maalampi and M. Raidal, {\sl Nucl. Phys.}{\bf B420}, 
449 (1994); C.S.Aulakh,A.Melfo and G.Senjanovi{\'c},
{\sl Phys.Rev.}{\bf D57},4174 (1998); G. Barenboim and N. Rius,
{\sl Phys. Rev.}{\bf D58}, 065010, (1998); N. Setzer and S. Spinner,
{\sl Phys. Rev.} {\bf D71}, 115010 (2005).

\bibitem {doublet}K. S. Babu.B. Dutta and R.N. Mohapatra, {\sl Phys.Rev.}
{\bf D65}, 016005, (2002).

\bibitem{melfo2}C. S. Aulakh, A. Melfo and Goran Senjanovi\'c ,  {\sl Phys.Rev.}{\bf D57}, 4174 (1998).

\bibitem{m86} R. N. Mohapatra,  Phys. Rev. {\bf D34} 3457 (1986);
A. Font, L. E. Ib\'a\~nez and F. Quevedo,  Phys. Lett. {\bf B228} 79 (1989); 
L. Ib\'a\~nez and G. Ross,  Phys. Lett. {\bf B260} 291 (1991); 
S. P. Martin,  Phys. Rev. {\bf D 46} 2769 (1992).


\bibitem{glu} P. Fayet,  {\sl Phys. Lett.} {\bf B78}, 417 (1978).

\bibitem{cmmc} C.M. Maekawa and M. C. Rodriguez, JHEP \textbf{04}, 031
(2006).

\bibitem{kraml} S. Kraml, hep-ph/9903257;
 J. Rosiek, Phys. Rev. {\bf D41}, 3464 (1990);

\bibitem{sps1}B.C. Allanach {\it et al}, {\sl Eur.Phys.J.}{\bf C25}, 113 (2002).

\bibitem{sps2}Nabil Ghodbane, Hans-Ulrich Martyn, hep-ph/0201233.

\bibitem{sps} http://spa.desy.de/spa/

\bibitem{Dawson}
  S.~Dawson, E.~Eichten and C.~Quigg,
  {\sl Phys. Rev.}  {\bf D31}, 1581 (1985).


\bibitem{Zerwas} W. Beenakker, R. H\"opker, M. Spira and P.M. Zerwas, {\sl Nucl. Phys.}{\bf B492}, 51 (1997).

\bibitem{plehn}
W.~Beenakker, M.~Kr\"amer, T.~Plehn, M.~Spira and P.~M.~Zerwas,
{\sl Nucl. Phys.} {\bf B515}, 3 (1998).


\bibitem{Pumplin:2002vw}
  J.~Pumplin, D.~R.~Stump, J.~Huston, H.~L.~Lai, P.~Nadolsky and W.~K.~Tung,
  JHEP {\bf 0207}, 012 (2002).

\end{thebibliography}
\end{document}